\documentclass[a4paper,twocolumn,aps,prb,superscriptaddress,showpacs,floatfix]{revtex4}
\usepackage{amssymb,amsmath,stmaryrd}
\usepackage{times}

\usepackage{float}
\usepackage{amsmath}
\usepackage{pst-node}
\usepackage{graphicx}
\usepackage{latexsym}
\usepackage{amsfonts}
\usepackage{amssymb}
\usepackage[nouppercase]{scrpage2}
\usepackage{makeidx}
\usepackage{bbm}
\usepackage{cancel}
\usepackage{amsthm}

\newcommand{\be}{\begin{equation}}
\newcommand{\ee}{\end{equation}}
\newcommand{\bea}{\begin{eqnarray}}
\newcommand{\eea}{\end{eqnarray}}

\def\a{\alpha}
\def\b{\beta}
\def\g{\gamma}

\def\d{\delta}
\def\D{\Delta}
\def\e{\epsilon}

\def\h{\eta}
\def\th{\theta}

\def\l{\lambda}

\def\m{\mu}
\def\n{\nu}

\def\p{\pi}

\def\r{\rho}
\def\s{\sigma}
\def\S{\Sigma}

\def\vf{\varphi}
\def\F{\Phi}

\def\w{\omega}
\def\W{\Omega}

\def\Q{\Psi}

\def\bgh{\mbox{\boldmath $\eta$}}

\def\blk{{\mathbf k}}

\def\blp{{\mathbf p}}
\def\blq{{\mathbf q}}
\def\blr{{\mathbf r}}

\def\blx{{\mathbf x}}

\def\blA{{\mathbf A}}

\def\callL{\mbox{$\mathcal{L}$}}

\def\callO{\mbox{$\mathcal{O}$}}

\def\callS{\mbox{$\mathcal{S}$}}
\def\callT{\mbox{$\mathcal{T}$}}

\def\ra{\rightarrow}

\def\de{\partial}
\def\iif{\infty}
\def\bra{\langle}
\def\ket{\rangle}

\def\Re{{\rm Re}}
\def\Im{{\rm Im}}

\def\1op{\hat{\mathbbm{1}}}
\def\1{\mathbbm{1}}
\def\nn{\nonumber}

\makeatletter
\newcommand{\subalign}[1]{%
  \vcenter{%
    \Let@ \restore@math@cr \default@tag
    \baselineskip\fontdimen10 \scriptfont\tw@
    \advance\baselineskip\fontdimen12 \scriptfont\tw@
    \lineskip\thr@@\fontdimen8 \scriptfont\thr@@
    \lineskiplimit\lineskip
    \ialign{\hfil$\m@th\scriptstyle##$&$\m@th\scriptstyle{}##$\crcr
      #1\crcr
    }%
  }
}
\makeatother

\begin{document}

\title{First-principles approach to excitons in time-resolved and
angle-resolved photoemission spectra}

\author{E. Perfetto}
\affiliation{Dipartimento di Fisica and European Theoretical Spectroscopy Facility (ETSF), 
Universit\`{a} di Roma Tor Vergata,
Via della Ricerca Scientifica 1, 00133 Rome, Italy}
\affiliation{Istituto di Struttura della Materia of the National Research 
Council, Via Salaria Km 29.3, I-00016 Montelibretti, Italy; and European Theoretical Spectroscopy Facility (ETSF)}
\author{D. Sangalli}
\affiliation{Istituto di Struttura della Materia of the National Research 
Council, Via Salaria Km 29.3, I-00016 Montelibretti, Italy; and European Theoretical Spectroscopy Facility (ETSF)}
\author{A. Marini}
\affiliation{Istituto di Struttura della Materia of the National Research 
Council, Via Salaria Km 29.3, I-00016 Montelibretti, Italy; and European Theoretical Spectroscopy Facility (ETSF)}
\author{G. Stefanucci}
\affiliation{Dipartimento di Fisica and European Theoretical Spectroscopy Facility (ETSF), 
Universit\`{a} di Roma Tor Vergata,
Via della Ricerca Scientifica 1, 00133 Rome, Italy}
\affiliation{INFN, Sezione di Roma Tor Vergata,
Via della Ricerca Scientifica 1, 00133 Roma, Italy}


\begin{abstract}
We show that any {\em quasi-particle} or GW approximation to the 
self-energy does not capture excitonic features in time-resolved 
(TR) photoemission spectroscopy. In this work
we put forward a first-principles approach and propose a feasible 
diagrammatic approximation to solve this problem. We also 
derive an alternative formula for the TR photocurrent 
which involves a single time-integral of the lesser Green's function. 
The diagrammatic approximation applies to the {\em relaxed} regime 
characterized by the presence of quasi-stationary  excitons 
and vanishing polarization.
The main distinctive feature of the theory is that the diagrams must 
be evaluated using {\em excited} Green's functions. 
As this is not standard the analytic derivation is presented in detail. 
The final result is an expression for the lesser Green's function in 
terms of quantities that can all be calculated {\em ab initio}.
The validity of the proposed theory is illustrated in a 
one-dimensional model system with a direct gap. We discuss 
possible 
scenarios and highlight some universal features of the exciton 
peaks. Our results indicate
that the exciton dispersion can be observed in TR {\em and} angle-resolved 
photoemission.
\end{abstract}


\maketitle

\section{Introduction}
\label{intro}

Time-resolved (TR) and angle-resolved 
photoemission (PE) spectroscopy  has established as a 
powerful experimental technique to monitor the femtosecond dynamics 
of electronic excitations in solid state physics. Applications cover  
the ultrafast dynamics in image potential 
states,\cite{GHHRS.1995,FW.1995,EBCFGH.2004,VMR.2004,GNEWW.2015}
electron relaxation in metals,\cite{FSTB.1992,PO.1997,SAEMMCG.2004,LLBSGW.2004} 
semiconductors~\cite{WKFR.2004,SS.2009,Netal.2014,WZR.2015} and 
more recently topological 
insulators,~\cite{RGKCH.2014,SYACFKS.2012,WHSSGLJG.2012,CRCZGBBKGP.2012,NOHFMEAABEC.2014,BVLNL.2014}  
charge transfer processes at 
solid state interfaces~\cite{GWLMGH.1998,VZ.1999,MYTZ.2008,ZYM.2009,VMT.2011} 
and in adsorbate on 
surfaces~\cite{HBSW.1997,Setal.2002,ZGW.2002,OLP.2004,MSMYLBAMK.2008,TY.2005,Aetal.2003},
and the formation and dynamics of  
excitons.~\cite{WKFR.2004,SS.2009,Z.2015,SRSW.1980,VBBT.2009,DWMRWS.2014} The 
theoretical description of  excitons 
constitutes the main focus of the present work. 
\begin{figure}[tbp]  
\includegraphics*[width=.46\textwidth]{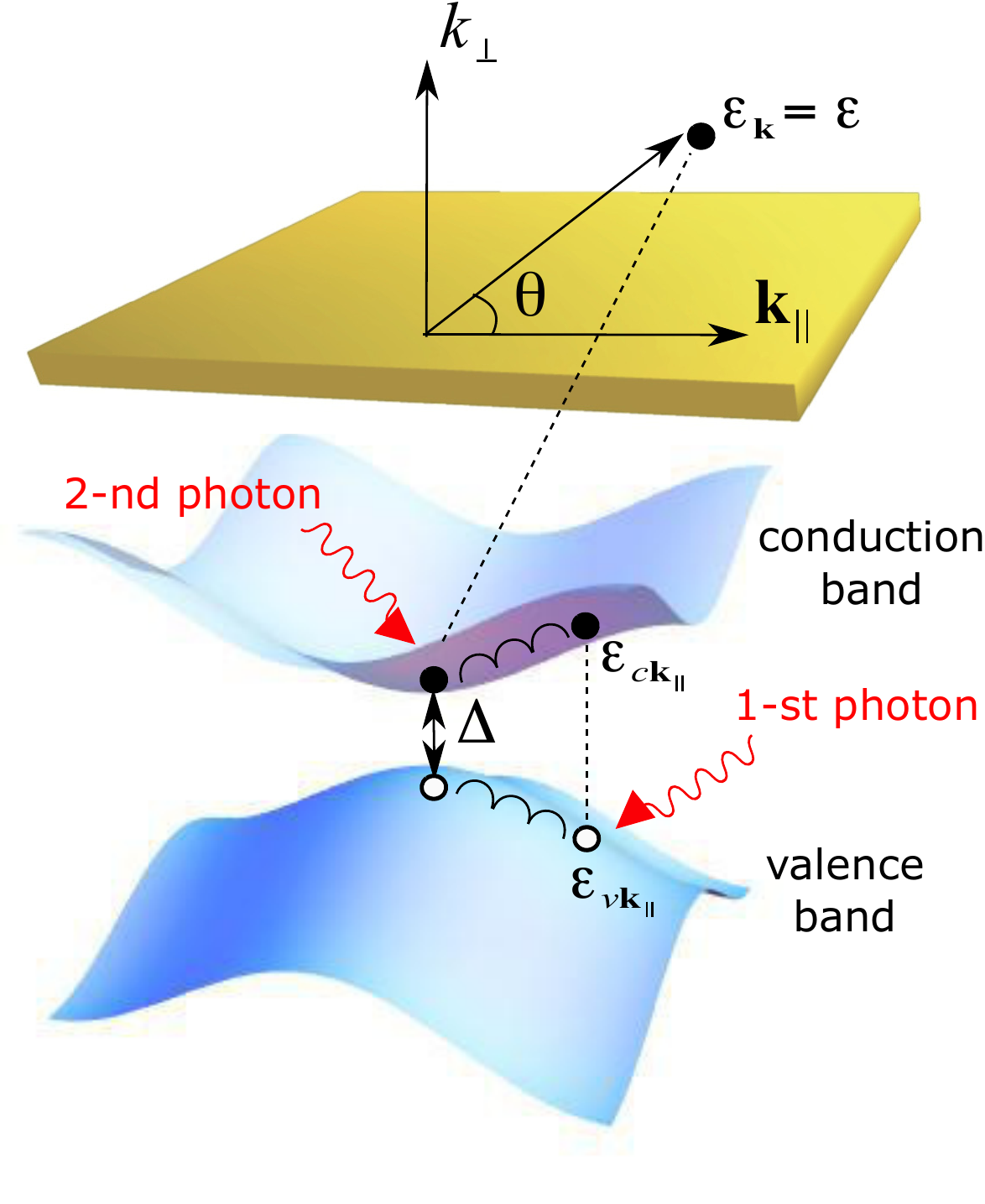}  
\caption{Schematic description of a TR and angle resolved 
PE experiment.}  
\label{expfig}  
\end{figure}  

In TR-PE  experiments on semiconductors or insulators a {\em pump} pulse excites 
electrons from the valence band to the conduction band. 
During the action of the pump the system coherently oscillates between the 
ground state and the dipole-allowed excited states giving rise to a 
finite polarization and hence to the emission of electromagnetic waves. 
Due to the Coulomb attraction between the conduction electrons and the 
valence holes the excited states may contain bound electron-hole 
({\em eh}) pairs or
{\em excitons}. If so then the lowest frequency of the 
time-dependent polarization (or, equivalently, the onset of the 
photoabsorption spectrum) reduces by an amount given by the exciton 
binding energy. In this oscillatory regime the system is not in an 
eigenstate and we say that it contains {\em virtual} 
excitons.\cite{KKKG.2006} After the pump has died off electrons (holes) remain trapped in the 
conduction (valence) band and relax toward the conduction band 
minimum (valence band maximum) because of inelastic 
scattering, see Fig.~\ref{expfig} for a schematic illustration. 
The relaxation process typically occurs on a femtosecond 
timescale\cite{SM.2002,Betal.1995,BC.1997,sm.2015} 
and the resulting quasi-stationary state is an {\em eh}
liquid containing {\em real} excitons, i.e., stationary 
bound {\em eh} pairs.\cite{KKKG.2006} In this regime we do not have a 
superposition of ground state and excited states but an {\em 
admixture} of them (hence the polarization vanishes).
The photocurrent of a TR-PE  experiment is generated by a  {\em probe} 
pulse which impinges the system in this quasi-stationary state and causes the emission of 
electrons from the conduction band. 
Like virtual excitons have an effect  on the photoabsorption spectrum so 
real excitons leave clear fingerprints on the TR-PE spectrum.

The photoabsorption spectrum is proportional to the 
polarization which, in turn, can be calculated from the 
Fourier transform of the time-dependent electron density $n(\blr,t)$. In the Green's function language 
$n(\blr,t)$ is given by the off-diagonal (in the basis of 
Bloch states) {\em equal-time} lesser Green's function $G^{<}(t,t)$. 
The effects of virtual excitons are therefore encoded in this 
quantity. It is well known that virtual excitons emerge
already when the equation of motion for $G^{<}(t,t)$ is solved
at the Hartree-Fock (HF) level. 
To lowest order in the perturbing 
field the HF  $G^{<}(t,t)$ can alternatively be obtained 
from the {\em equilibrium} density response function 
which solves the Bethe-Salpeter equation (BSE) with HF 
kernel.\cite{Fetterbook,SvLbook,ullrichbooh} In 
more refined state-of-the-art first-principles calculations the HF 
kernel is replaced by a Hartree plus 
screened exchange\cite{H.1965} (HSEX)
kernel.\cite{s.1988,SMH.1982,ORR.2002,ARDSO.1998,BSB.1998,RL.1998,PPHS.2011,AGM.2011,PSMS.2015}
The theory of excitons in photoabsorption spectra is today very well 
established.

Conceptually different is the TR-PE spectrum since it is proportional to 
the probability of finding an electron with a certain momentum and 
energy. In the Green's function language this probability is 
given by the Fourier transform of the diagonal (in the basis of 
Bloch states) lesser Green's function $G^{<}(t-t')=G^{<}(t,t')$ (the 
dependence on the time-difference only is a consequence of the 
quasi-stationarity of the system). It could be tempting to calculate 
the quasi-stationary $G^{<}(t-t')$ in the HSEX approximation since the 
{\em equal-time} HSEX 
$G^{<}(t,t)$ contains the physics of virtual excitons.
However, we anticipate that real excitons do not emerge from 
HSEX.
 In fact, at present there exist no first-principles diagrammatic
approach to calculate the impact of real excitons on the TR-PE 
spectrum. The purpose of the present work is to fill this gap. 

The paper is organized as follows. In Section~\ref{picturesec} we 
briefly discuss a simple  
picture of the exciton problem in TR-PE spectroscopy.
In 
Section~\ref{NEQphotocurrentsec} we derive a general formula of the 
TR photocurrent valid for arbitrary
intensities and shapes of the probe field and  
involving a single time-integral of the 
lesser Green's function. The inadequacy of the HF, HSEX and GW
approximations to $G^{<}(t-t')$
is illustrated in  Section~\ref{failuresec}.
In Section~\ref{diagtreatsec} we  
identify the relevant diagrams to calculate the dressed
Green's function. We show that the (self-energy) vertex 
should satisfy a {\em nonequilibrium} Bethe-Salpeter equation (BSE) with a 
HSEX kernel evaluated at {\em excited} quasiparticle 
 ({\em qp})
Green's functions. In 
Section~\ref{BSELsec} we generalize the solution of the nonequilibrium 
BSE of Ref.~\onlinecite{SRFHB.2011} to arbitrary momenta  and show how 
to extract the 
lesser and greater component of the {\em eh} propagator. 
This part of the theory is also useful to calculate 
photoluminescence spectra.\cite{HGB.2000} From the lesser and greater {\em eh} 
propagators we 
construct  the (self-energy) vertex and subsequently the 
spectral function, see Section~\ref{Sigma-G-sec}. Taking into account the quasi-stationarity of the system we 
finally obtain a simple and intuitive expression for the (dressed) lesser 
Green's function. The proposed  
treatment is benchmarked in a minimal model with only one valence band and 
one conduction band. For the case of a single {\em eh} pair 
the model can be solved analytically and our
diagrammatic treatment is shown to be {\em exact}, see Section~\ref{analytic-solution-sec}.
In Section~\ref{finite-t-sec} we consider 
a finite {\em eh} density, discuss possible scenarios and highight some 
universal features of the excitonic features. A summary of the method 
and the main conclusions are drawn in Section~\ref{conclusions-sec}.
 
\section{A simple physical picture}
\label{picturesec}

Let us briefly illustrate a simple 
physical picture of TR-PE in systems with real 
excitons.\cite{Z.2015} After absorption of a pump 
photon the system makes a transition, from the ground state of energy $E_{g}$ to 
an excited state of energy $E$ characterized by one electron in the conduction 
band. Subsequently, the conduction electron absorbes 
a (probe) photon of energy $\w_{0}$ and it is
expelled  as a photoelectron of momentum $\blk$ and energy 
$\e_{\blk}>0$ (we set the continuum threshold to zero). Energy conservation and conservation of 
the momentum parallel to the surface imply that  
$\w_{0}+E=E^{-}_{\blk_{\parallel}}+\e_{\blk}$, where $E^{-}_{\blk_{\parallel}}$ is the energy of 
the original system {\em without} a valence electron of momentum 
$\blk_{\parallel}$ and energy $\e_{v\blk_{\parallel}}$. 
Approximating $E^{-}_{\blk_{\parallel}}\simeq 
E_{g}-\e_{v\blk_{\parallel}}$ one finds 
the
{\em momentum resolved} photocurrent 
\be
I(\blk)\propto 
\d(\w_{0}+E-E_{g}+\e_{v\blk_{\parallel}}-\e_{\blk}),
\label{heupc}
\ee
from which it follows that the
{\em energy-resolved} photocurrent {\em perpendicular} to the surface 
is 
\be
I(\e)\propto \d(\w_{0}+E-E_{g}+\e_{v0}-\e).
\ee
If the {\em eh} pair of the excited state does not bound then 
$E-E_{g}$ is no smaller than the optical gap $\D$ and the 
photocurrent is nonvanishing  for 
$\e>\w_{0}+\D+\e_{v0}$. If, on the other hand, the {\em eh} pair 
bounds then the lowest excited state splits off from the continuum by 
an amount equal to
the exciton binding energy $b_{X}$ and the photocurrent is nonvanishing 
also at the discrete energy 
values $\e=\w_{0}+\D+\e_{v0}-b_{X}$. Thus, the formation of an exciton 
manifests as a photocurrent peak 
below the onset of the continuum.

Although this picture captures the qualitative aspects of the 
problem, it lacks of a quantitative description of the phenomenon. 
In reality, after the action of the pump pulse the system is not in a pure state 
characterized by a single {\em eh} pair but in 
an admixture of excited states with a certain distribution of {\em eh} 
pairs and the exciton binding 
energy depends on this distribution in a far from obvious manner.
The above 
picture is also inadequate to determine the proportionality constant in 
Eq.~(\ref{heupc}), thus preventing a quantitative comparison with the 
experiment. 

The failure of the HF or HSEX (or any other {\em qp} 
for that matter) approximation is also evident. 
Due to Coulomb attraction with the valence hole 
the bare conduction electron splits into a conduction 
{\em qp} of roughly the same energy and a {\em qp} bound to the 
valence hole. In other words every bare electron, characterized by a 
well defined energy, is transformed into two {\em qp}'s of different 
energies. By construction a {\em qp} approximation assigns a single energy to every {\em 
qp} and it is therefore inadequate to study real excitons in TR-PE.
A more technical discussion of this point can be found in 
Section~\ref{failuresec} while in Section~\ref{diagtreatsec}
we propose a diagrammatic solution to the
problem. Preliminarly, however, we derive a formula which relates the 
TR photocurrent to the lesser Green's function.

\section{Nonequilibrium Photocurrent}
\label{NEQphotocurrentsec}

In this Section we derive and discuss the formula for the 
time-dependent photocurrent induced by a laser pulse impinging on a 
solid out of equilibrium. 
By definition 
the photocurrent of electrons with momentum $\blk=(\blk_{\parallel},k_{\perp})$ is 
given by the rate of change of the occupation of the 
time-reversed low-energy electron-diffraction (LEED) 
state\cite{SKRL.1993,SDCZB.2008,BRPME.2015} with  momentum $\blk$, i.e.,
\bea
I(\blk,t)&\equiv& \frac{d}{dt}\bra 
\hat{f}^{\dag}_{H\blk}(t)\hat{f}_{H\blk}(t)\ket
\nn\\
&=&-i\frac{d}{dt}G^{<}_{f\!f,\blk}(t,t),
\label{photocurr}
\eea
where $\hat{f}_{\blk}$ annihilates an electron in the 
LEED state of momementum $\blk$ and the subindex $H$ signifies that 
operators evolve according to the Heisenberg picture in the presence 
of the pump and probe fields. In the second line of Eq.~(\ref{photocurr}) 
appears the lesser component of the free-electron Green's function 
which is defined according to~\cite{SvLbook} 
\be
G_{f\!f,\blk}(z,z')\equiv \frac{1}{i}\bra\callT\left\{\hat{f}_{H\blk}(z)
\hat{f}^{\dag}_{H\blk}(z')\right\}\ket,
\label{keldyshg}
\ee
where $z$ and $z'$ are times on the Keldysh contour and $\callT$ is 
the contour ordering operator.
Denoting by $\e_{f\blk}=k^{2}/2>0$ the free-electron energy, the 
LEED states are linear combination of Bloch states with 
energy $\e_{f\blk}$.\cite{SDCZB.2008} We refer to 
Refs.~\onlinecite{CI.1999,SDCT.2004} for the description of an 
efficient numerical algorithm to calculate these states. 
We work in the dipole approximation (which is accurate for photon 
energies below 10 keV) and consider the vector potential 
of the probe field
$\blA(t)=\bgh\, a(t)$ parallel to some unit vector $\bgh$. As we are 
interested in the photocurrent generated by a pulse the function 
$a(t)$ vanishes for $t\to\pm\iif$.
Let $D_{\n\blk}$ 
be the matrix element of $(\blp\cdot\bgh)/c$ 
between a LEED state of momentum 
$\blk=(\blk_{\parallel},k_{\perp})$ and a bound Bloch state (of energy 
below zero) with band-index $\n$ and parallel momentum 
$\blk_{\parallel}$ (parallel momentum is conserved). 
Neglecting the Coulomb interaction between LEED electrons
and bound electrons in the solid, the equations of motion for 
$G_{f\!f,\blk}$ read
\bea
\left[i\frac{d}{dz}-\e_{f\blk}\right]\!G_{f\!f,\blk}(z,z')-
\sum_{\n}D^{\ast}_{\n\blk}a(z)G_{\n f,\blk}(z,z')
\nn\\
=\d(z,z'),
\label{eom1}
\eea
\bea
\left[-i\frac{d}{dz'}-\e_{f\blk}\right]\!G_{f\!f,\blk}(z,z')-
\sum_{\n}D_{\n\blk}a(z')G_{f\n,\blk}(z,z')
\nn\\=\d(z,z'),
\label{eom2}
\eea
where $G_{f\n,\blk}(z,z')$ and 
$G_{\n f,\blk}(z,z')$ are defined {\em mutatis mutandis} as in Eq.~(\ref{keldyshg}).
Equations~(\ref{eom1}-\ref{eom2}) and all subsequent equations of motion
have to be solved with Kubo-Martin-Schwinger boundary conditions.\cite{SvLbook}
Setting $z=t_{-}$ and $z'=t_{+}$  and subtracting Eq.~(\ref{eom2}) 
from Eq.~(\ref{eom1}) we find
\be
i\frac{d}{dt}G^{<}_{f\!f,\blk}(t,t)=-2\,\Re\left[
\sum_{\n}D_{\n\blk}a(t)G^{<}_{f\n,\blk}(t,t)\right].
\label{g<ff}
\ee
We can express the right hand side of Eq.~(\ref{g<ff}) in terms of the Green's 
function $G_{\n'\n,\blk_{\parallel}}(z,z')\equiv \frac{1}{i}\bra\callT\left\{\hat{c}_{\n \blk_{\parallel}}(z)
\hat{c}^{\dag}_{\n' \blk_{\parallel}}(z')\right\}\ket$ with both 
indices in the bound Bloch sector. The equation of 
motion for $G_{f\n,\blk}$ reads
\be
\left[i\frac{d}{dz}-\e_{f\blk}\right]G_{f\n,\blk}(z,z')-
\sum_{\n'}D^{\ast}_{\n'\blk}a(z)G_{\n'\n,\blk_{\parallel}}(z,z')=0.
\label{eom3}
\ee
If we define the unperturbed (probe-free) Green's function as the solution of
\be
\left[i\frac{d}{dz}-\e_{f\blk}\right]g_{f\!f,\blk}(z,z')=\d(z,z'),
\nn
\ee
then Eq.~(\ref{eom3}) can be solved for $G_{f\n,\blk}$ yielding
\be
G_{f\n,\blk}(z,z')=\sum_{\n'}\int \!\!d\bar{z}\,g_{f\!f,\blk}(z,\bar{z})\,
D^{\ast}_{\n'\blk}a(\bar{z})G_{\n'\n,\blk_{\parallel}}(\bar{z},z').
\nn
\ee
Substituting this result into Eq.~(\ref{g<ff}) we see that it is  
convenient to define the {\em embedding self-energy}
\be
\S_{\n\n',\blk}^{\rm emb}(z,z')\equiv
D_{\n\blk}a(z)g_{f\!f,\blk}(z,z')a(z')D^{\ast}_{\n'\blk}.
\label{embse}
\ee
The embedding self-energy accounts for the fact that electrons can 
escape from the solid.\cite{PUvLS.2015,SBP.2016,PUvLS.2016} A similar quantity is 
used in the context of quantum transport where the electrons of a molecular 
junction can move in and out of the junction by tunneling from and to the 
leads.\cite{MW.1992,JWM.1994,SA.2004} The complex 
absorbing potential in quantum mechanics can be seen as a time-local 
approximation to $\S^{\rm emb}$. It is worth noticing that the 
embedding self-energy is independent of the 
electron-electron and electron-phonon interactions and it is completely determined by
the matrix elements $D_{\n\blk}$ and by the pulse shape $a(t)$.

Using the Langreth 
rules~\cite{SvLbook} and taking into account that $\S_{\n\n',\blk}^{\rm emb,<}\propto 
g^{<}_{f\!f,\blk}\propto f(\e_{f\blk})=0$ since there are no LEED 
electrons in the initial state (here $f(\e)$ is the Fermi function),
we can rewrite Eq.~(\ref{g<ff}) as
\be
I(\blk,t)=2\sum_{\n\n'}\int \!\!d\bar{t}\;\Re\left[
\S_{\n\n',\blk}^{\rm emb,R}(t,\bar{t})G_{\n'\n,\blk_{\parallel}}^{<}(\bar{t},t)
\right],
\label{photocurr2}
\ee
where
\be
\S_{\n\n',\blk}^{\rm emb,R}(t,\bar{t})=-i\th(t-\bar{t})
D_{\n\blk}D^{\ast}_{\n'\blk}
a(t)a(\bar{t})e^{-i\e_{f\blk}(t-\bar{t})}.
\nn
\ee
This is our formula for the {\em time-dependent photocurrent} 
and it constitues the main result of this Section. The formula is valid for systems 
in arbitrary nonequilibrium states and for any 
temporal shape and {\em intensity} of the probe field, the 
only approximation being that LEED electrons do not interact with 
bound electrons.
We observe that 
Eq.~(\ref{photocurr2}) reduces to the formula  derived in Ref.~\onlinecite{FKP.2009} 
provided that one approximates $\frac{d}{dt}\bra 
\hat{f}^{\dag}_{H\blk}(t)\hat{f}_{H\blk}(t)\ket\simeq |\blk|
\bra\hat{f}^{\dag}_{H\blk}(t)\hat{f}_{H\blk}(t)\ket$ and 
discards the effect of the probe field on $G_{\n'\n,\blk_{\parallel}}$. 
A practical numerical advantage of Eq.~(\ref{photocurr2}) 
is that it contains a single time integral.

To make contact with the discussion of the introductory Section 
we consider the special 
case of a system left 
in a stationary excited state after the action of the pump pulse\cite{note}
and take a probe pulse sharply peaked at frequency $\w_{0}$, i.e.,
$a(t)=\th(t)\left(a_{0}e^{i\w_{0} t}+c.c.\right)$.  
If we are interested in the photocurrent for $t\to\infty$ only the 
terms depending on the time-difference contribute to the embedding 
self-energy.  If we further assume (as in the introductory Section) 
that electrons are expelled from the conduction band $\n=c$ then we 
can limit the sum in Eq.~(\ref{photocurr2}) to $\n=\n'=c$ using
\bea
\S_{cc,\blk}^{\rm emb,R}(t,\bar{t})\!\!&=&\!\!-i\th(t-\bar{t})
|a_{0}D_{c\blk}|^{2}e^{-i\e_{f\blk}(t-\bar{t})}
\nn\\
\!\!&\times&\!\!
(e^{i\w_{0} (t-\bar{t})}+c.c.).
\label{monosigma2}
\eea
To lowest order in the probe field $G_{cc,\blk_{\parallel}}^{<}$ 
depends on the time difference only (the system is in 
a stationary state). 
Inserting Eq.~(\ref{monosigma2}) into Eq.~(\ref{photocurr2}) we then find
\bea
I(\blk,t)&=&-2|a_{0}D_{c\blk}|^{2}\int\frac{d\w}{2\p}\,iG_{cc,\blk_{\parallel}}^{<}(\w)
\nn\\
&\times&
\Re\left[\int_{0}^{t} 
\!\!d\bar{t}\;\left(e^{-i\W_{-}(t-\bar{t})}+e^{-i\W_{+}(t-\bar{t})}\right)\right],
\quad
\nn
\eea
where we used that $iG_{cc,\blk_{\parallel}}^{<}(\w)$ is real and we defined 
$\W_{\pm}=\e_{f\blk}\pm\w_{0}-\w$. Performing the time integral and 
taking into account that $\lim_{t\ra\iif}\frac{\sin\W 
t}{\W}=\p\d(\W)$,
the long-time limit of the photocurrent is given by
\bea
I(\blk)\!\!&\equiv&\!\! \lim_{t\ra\iif}I(\blk,t)
\nn\\
\!\!&=&\!\!
-i|a_{0}D_{\blk}|^{2}\!\left[
G_{cc,\blk_{\parallel}}^{<}(\e_{f\blk}-\w_{0})+G_{cc,\blk_{\parallel}}^{<}(\e_{f\blk}+\w_{0})\right].
\nn\\
\label{photocurr3}
\eea
Comparing this result with Eq.~(\ref{heupc}) we see that  a 
proper selection of Feynman diagrams evaluated with an {\em excited} 
{\em qp} 
Green's function are required to capture 
excitonic features in the energy-resolved photocurrent. In fact, 
$G_{cc,\blk_{\parallel}}^{<}(\w)$  is nonvanishing at 
the removal energies of the {\em excited} solid. 
In the next two Sections we develop a diagrammatic treatment to tackle 
this problem.

\section{Failure of quasi-particle and GW approximations}
\label{failuresec}

In order to avoid the numerically expensive 
implementation of the two-times Kadanoff-Baym 
equations\cite{SvLbook,KBbook,KB.2000,DvL.2007,MSSvL.2009,BB.2013,PvFVA.2009,SBP.2016,SB.2016} 
the lesser Green's function is usually calculated from the 
Generalized Kadanoff-Baym 
Ansatz\cite{lsv.1986,bonitz.book,anttibook,BSH.1999,HB.1996,M.2013,LPUvLS.2014}
(GKBA)
\be
G^{<}(t,t')=iG^{\rm R}(t,t')G^{<}(t',t')-iG^{<}(t,t)G^{\rm A}(t,t'),
\ee
where $G^{\rm R}(t,t')=[G^{\rm A}(t',t)]^{\dag}$ is the retarded 
Green's function in some {\em qp} approximation, e.g., 
HF or HSEX. It is well established that the {\em equal-time} HSEX $G^{<}$ 
accurately describes virtual excitons in photoabsortion (the photoabsorption spectrum is 
proportional to $\int dt \,e^{i\w t}G^{<}(t,t)$).\cite{AGM.2011} 
Real excitons, 
however, arise from the Fourier transform of $G^{<}(t,t')$ with 
respect to the {\em relative-time} $(t-t')$; therefore real excitons
hide in $G^{\rm R}(t,t')$ and not in $G^{<}(t,t)$. 
In any {\em qp} approximation $G^{\rm R}(t,t')$ is a {\em single} oscillatory 
exponential with frequency given by the {\em qp} energy. Thus, the 
Fourier transform $G^{<}(\w)$ 
is peaked {\em only} at the {\em qp} energy and does not contain  
information on the exciton peak. The very same 
approximation which accurately describes virtual excitons (in 
photoabsorption) fails to describe real excitons (in TR-PE). 
The situation does not improve at the GW level.
In fact,  in insulators and semiconductors the 
main effect of the GW self-energy  is to renormalize the {\em qp}  
energies. Dynamical effects (due to 
the dependence on frequency) appear at very high energy 
and are associated to plasmonic excitations, not to excitons. 
Hence, the retarded Green's function in the GW approximation 
maintains a {\em qp} 
character. 

To make progress we must abandon the {\em qp} approximation 
and calculate $G^{\rm R}$ using a  
many-body self-energy $\S$ with {\em vertex corrections}. We emphasize that $\S$ 
is distinct from the embedding self-energy defined in 
Eq.~(\ref{embse}): the former is a functional of the Green's function 
and Coulomb 
interaction whereas the latter is an explicit 
functional of the probe pulse. Hence $\S$ is nonvanishing even without a 
probe whereas $\S^{\rm emb}$ is nonvanishing even without the Coulomb 
interaction.

\section{Diagrammatic treatment}
\label{diagtreatsec}

To find the most relevant many-body self-energy diagrams we 
argue as follows. In a metal the plasmon peak in 
photoabsorption is captured by a two-particle Green's function $G_{2}$
evaluated from the 
Bethe-Salpeter equation (BSE) with Hartree kernel $K_{\rm H}=-\d\S_{\rm H}/\d G$. 
However, in PE the plasmon peak does not emerge from a 
Green's function calculated with Hartree self-energy $\S_{\rm H}$.
Rather, the plasmon peak emerges from
the GW 
self-energy $\S_{\rm GW}\equiv -iv G_{2}G^{-1}$, where $v$ is the Coulomb interaction and 
$G_{2}$ is the two-particle Green's function which solves the BSE 
with kernel $K_{\rm H}$.
By analogy we expect that real excitons emerge 
from a self-energy $\S=-iv G_{2}G^{-1}$ where 
$G_{2}$ solves the BSE
with kernel $K_{\rm HSEX}=-\d\S_{\rm HSEX}/\d G$, $\S_{\rm HSEX}$ being 
the HSEX 
self-energy. In fact, 
this $G_{2}$ contains the $T$-matrix diagrams in the particle-hole 
sector which we know to describe the physics of excitons in 
photoabsorption. The twist 
with respect to the plasmon case is that in PE plasmons 
are seen also in equilibrium whereas excitons are not. As we 
shall see this aspect is not related to the selection of self-energy 
diagrams but to the {\em qp} Green's function chosen
to evaluate them.

\begin{figure}[tbp]  
\includegraphics*[width=.46\textwidth]{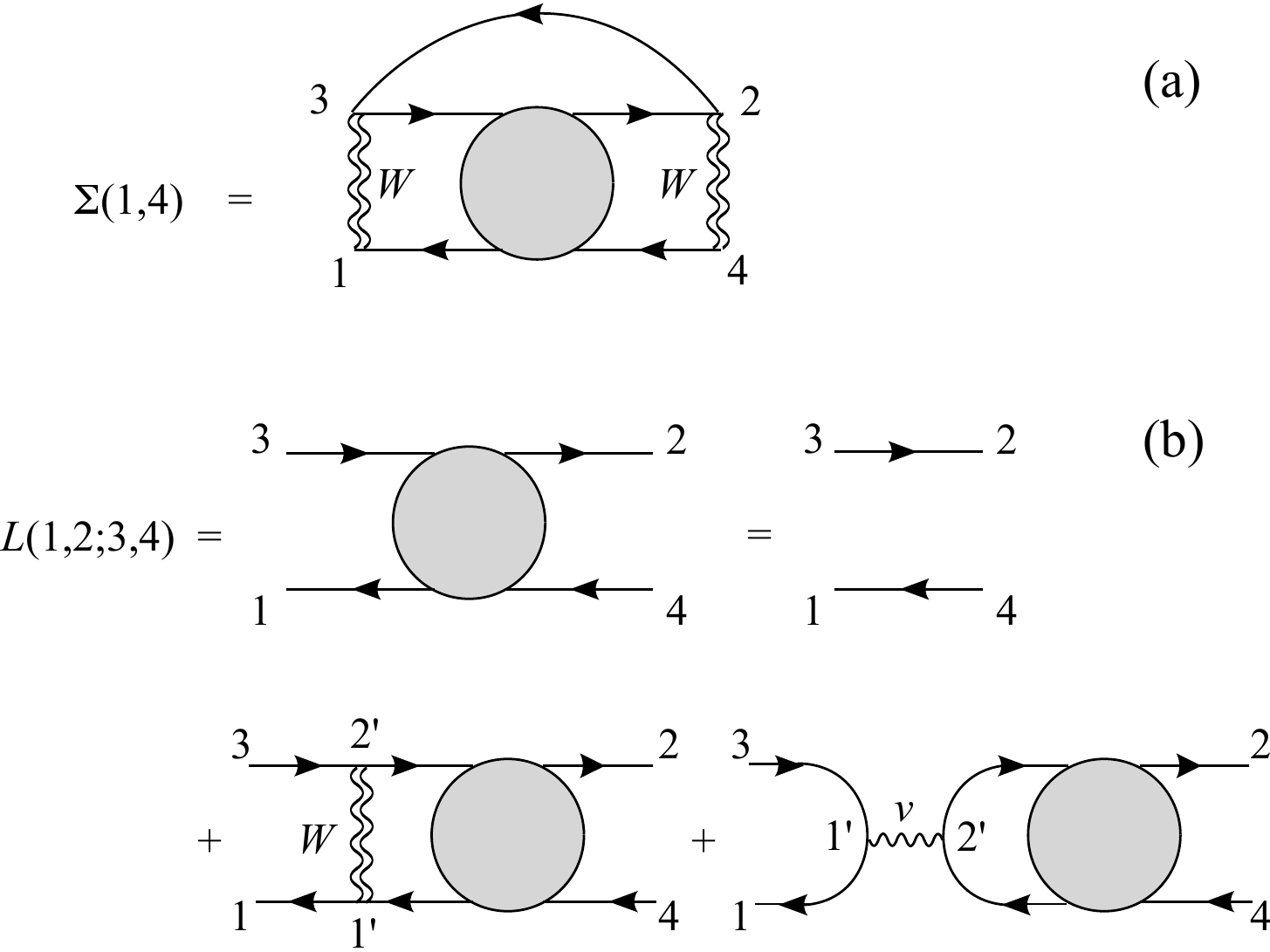}  
\caption{(a) Diagram for the self-energy. (b) Diagram for $L$. Wiggly 
lines denote the bare interaction $v$ and doubly wiggly lines denote the 
statically screened interaction $W$.}  
\label{sel}
\end{figure}  
On the basis of this discussion we propose to calculate the 
Green's function appearing in Eq.~(\ref{photocurr2}) using the self-energy in 
Fig.~\ref{sel}~(a) where the two-particle correlation function 
\be
L(1,2;3,4)\equiv -G_{2}(1,2;3,4)+G(1;3)G(2;4)
\nn
\ee
is given in Fig.~\ref{sel}~(b) and is evaluated
using {\em excited} {\em qp} Green's functions. 
The latter are calculated by 
performing numerical simulations of the dynamics of the system 
in the presence of the pump field.
This can be done fully {\em ab 
initio} using, e.g., the Yambo 
code~\cite{MHGV.2009} which implements a one-time Kadanoff-Baym evolution for the 
electronic 
populations.\cite{lsv.1986,bonitz.book,anttibook,BSH.1999,HB.1996,M.2013} Previous 
studies on bulk silicon~\cite{sm.2015,sm2.2015,SDCMCM.2016} 
have shown that the polarization dies off a few femtoseconds after the pump pulse 
due to inelastic scattering and that the pumped electrons reach a Fermi-Dirac 
distribution $f(\e)=1/(e^{\b(\e-\m)}+1)$ with band-dependent 
temperature $1/\b$ and chemical potential $\m$. Electron-hole 
recombination and hence relaxation toward the ground state
does instead occur on a picosecond time-scale. Thus, 
the solid is well described by an admixture of stationary excited 
states on the 
(femtosecond) time-scale  of the probe pulse.\cite{PSMS.2015} 
It is the purpose of this Section to develop a 
first-principles approach to nonequilibrium PE in such
regime.

\subsection{Excited two-particle correlation function}
\label{BSELsec}

As the screened interaction $W$ in Fig.~\ref{sel}~(a) is static, the 
vertices $(1,3)$ and $(2,4)$ have the same time argument. It is 
therefore sufficient to evaluate
\be
L_{\substack{\blx_{1}\blx_{3}\\ \blx_{2}\blx_{4}}}(z,z')\equiv
L(\blx_{1}z,\blx_{2}z';\blx_{3}z,\blx_{4}z'),
\nn
\ee
where $\blx=(\blr\s)$ is a collective index for the position and spin 
coordinate whereas $z$ is a contour time. The Green's function lines 
in Fig.~\ref{sel}~(b) describe {\em qp} propagators in some 
admixture of stationary excited states 
\be
g_{\blx_{1}\blx_{4}}(z,z')=\sum_{j}\vf_{i}(\blx_{1})\vf_{j}^{\ast}(\blx_{4})
g_{j}(z,z'),
\label{gexp}
\ee
where $\vf_{j}$ is the {\em qp} wavefunction and 
$j$ is a collective index for the band, spin and
momentum. Expanding $L$ according to
\be
L_{\substack{\blx_{1}\blx_{3}\\ \blx_{2}\blx_{4}}}(z,z')
=\sum_{\substack{ij\\mn}}
L_{\substack{ij\\mn}}(z,z')\vf_{i}(\blx_{1})
\vf_{j}^{\ast}(\blx_{3})\vf_{m}(\blx_{2})\vf_{n}^{\ast}(\blx_{4}),
\label{Lexp}
\ee
the BSE of Fig.~\ref{sel}(b) takes the form
\bea
L_{\subalign{&ij\\& mn}}(z,z')=\d_{in}\d_{jm}\,g_{i}(z,z')g_{j}(z',z)+
i\sum_{pq}\int d\bar{z}
\nn\\
\times g_{i}(z,\bar{z})g_{j}(\bar{z},z)
K_{\subalign{&ij\\& qp}} L_{\subalign{&pq\\& mn}}
(\bar{z},z'),
\label{Lgeneralbasis}
\eea
where $K_{\subalign{&ij\\& qp}}\equiv W_{iqjp}-v_{iqpj}$. Here  the 
four-index statically screened interaction is defined according to
\be
W_{ijmn}=\!\!\int\! \!d\blx_{1} d\blx_{2}\,\vf_{i}^{\ast}(\blx_{1})
\vf_{j}^{\ast}(\blx_{2})\vf_{m}(\blx_{2})
\vf_{n}(\blx_{1})W(\blx_{1},\blx_{2}).
\label{4indx}
\ee
The definition of the four-index bare interaction is analogous and 
is obtained by replacing $W$ with $v$ in 
Eq.~(\ref{4indx}).

To take advantage of the conservation of momentum we write every
label $i,j,\ldots$ in terms of a collective greek index 
that specifies band 
and spin, and a latin bold index that specifies the
value of the momentum, e.g., $i=\a\blk$, $j=\b\blp$, etc. 
Since we are describing  electrons bound to the solid
all momenta have vanishing 
component perpendicular to the surface. Momentum conservation implies 
that  the sum of the momenta of the 
indices $(i,q)$ in  $K_{\subalign{&ij\\& qp}}$ is the same as the sum of the momenta of the indices 
$(j,p)$. Therefore 
\be
K_{\subalign{&\m\blk+\blq,\n\blk\\& \a\blk''+\blq-\blq'',\b\blk''+\blq}}
=\d_{\blq\blq''} K^{\blq}_{\subalign{&\m\n\blk\\ &\a\b\blk''}},
\label{newKtensor}
\ee
which implicitly defines the tensor on the right hand side.
For a tensor $K$ with the property in Eq.~(\ref{newKtensor}) the 
solution of Eq.~(\ref{Lgeneralbasis}) is a tensor $L$ with the same property. Thus the 
BSE reduces to 
\bea
L^{\blq}_{\subalign{&\m\n\blk\\&\r\s\blk'}}(z,z')=
\d_{\m\s}\d_{\n\r}\d_{\blk\blk'}
g_{\m\blk+\blq}(z,z')g_{\n\blk}(z',z)
\nn\\+i\sum_{\a\b\blk''}
\int\! d\bar{z}\,
g_{\m\blk+\blq}(z,\bar{z})g_{\n\blk}(\bar{z},z)
K^{\blq}_{\subalign{&\m\n\blk\\ &\a\b\blk''}}
L^{\blq}_{\subalign{&\b\a\blk''\\&\r\s\blk'}}(\bar{z},z').
\quad
\label{Lblochbasis}
\eea
Introducing the superindices $I=(\m\n\blk)$, 
$J=(\s\r\blk')$ etc. and using the convention that lower 
superindices have swapped band-spin indices, e.g. $A_{\substack{I\\J}}=
A_{\substack{\m\n\blk\\\r\s\blk'}}$, we can rewrite 
Eq.~(\ref{Lblochbasis}) in the following compact form
\be
L^{\blq}_{\substack{I\\J}}(z,z')=\d_{\substack{I\\J}}
\ell^{\blq}_{I}(z,z')+i\sum_{M}\int\! d\bar{z}\, \ell^{\blq}_{I}(z,\bar{z})
K^{\blq}_{\substack{I\\ M}}
L^{\blq}_{\substack{M\\J}}(\bar{z},z'),
\label{Lsuperbasis}
\ee
where $\d_{\substack{I\\J}}=\d_{\substack{\m\n\blk\\\r\s\blk'}}\equiv
\d_{\m\s}\d_{\n\r}\d_{\blk\blk'}$ and
\be
\ell^{\blq}_{I}(z,z')=\ell^{\blq}_{\m\n\blk}(z,z')\equiv
g_{\m\blk+\blq}(z,z')g_{\n\blk}(z',z)
\nn
\ee
is the free {\em eh} propagator.
The Green's function $g$ is an excited {\em qp}  
Green's function and therefore the lesser and greater components are given by
\begin{subequations}
\begin{alignat}{2}    
&g^{<}_{\m\blk}(\w)=2\p i f_{\m\blk}\d(\w-\e_{\m\blk}),
\label{g0<}\\
&g^{>}_{\m\blk}(\w)=-2\p i \bar{f}_{\m\blk}\d(\w-\e_{\m\blk}),
\label{g0>}
\end{alignat}
\label{g0}
\end{subequations}
\vspace{-0.4cm}

\noindent
where $f_{\m\blk}$ is the {\em qp} occupation of level 
$\m\blk$ with energy $\e_{\m\blk}$ whereas 
$\bar{f}_{\m\blk}=1-f_{\m\blk}$. Since the solid is in an admixture 
of excited states the occupations do not follow a thermal 
distribution.
It is straightforward to extract the lesser/greater component 
of $\ell^{\blq}$:
\bea
\ell^{\blq,>}_{\m\n\blk}(\w)&=&\int\frac{d\w'}{2\p}
g^{>}_{\m\blk+\blq}(\w+\w')g^{<}_{\n\blk}(\w')
\nn\\
&=&2\p\bar{f}_{\m\blk+\blq}f_{\n\blk}\d(\w-\e_{\m\blk+\blq}+\e_{\n\blk}),\quad
\label{x>}
\eea
and similarly
\be
\ell^{\blq,<}_{\m\n\blk}(\w)=2\p 
f_{\m\blk+\blq}\bar{f}_{\n\blk}\d(\w-\e_{\m\blk+\blq}+\e_{\n\blk}).
\label{x<}
\ee
Therefore
\bea
\ell^{\blq,\rm R/A}_{\m\n\blk}(\w)&=&
i\int\frac{d\w'}{2\p}\frac{\ell^{\blq,>}_{\m\n\blk}(\w')-\ell^{\blq,<}_{\m\n\blk}(\w')}
{\w-\w'\pm i\h}
\nn\\
&=&i\frac{f_{\n\blk}-f_{\m\blk+\blq}}{\w-\e_{\m\blk+\blq}+\e_{\n\blk}\pm i\h}.
\label{xqra}
\eea
Again to keep the notation as light as possible we define
\be
f_{I}^{\blq}=f^{\blq}_{\m\n\blk}\equiv f_{\n\blk}-f_{\m\blk+\blq},
\label{superocc}
\ee
and
\be
\w^{\blq}_{I}=\w^{\blq}_{\m\n\blk}\equiv \e_{\m\blk+\blq}-\e_{\n\blk},
\nn
\ee
so that Eq.~(\ref{xqra}) takes the following compact form
\be
\ell^{\blq,\rm R/A}_{I}=i\frac{f_{I}^{\blq}}{\w-\w^{\blq}_{I}\pm i\h}.
\label{xrasuperindex}
\ee
We now proceed to the calculation of the various Keldysh components 
of $L$.

\subsubsection{Retarded component}
 Extracting the retarded component of Eq.~(\ref{Lsuperbasis}), 
Fourier transforming  
and using Eq.~(\ref{xrasuperindex}) we get
\be
(\w-\w^{\blq}_{I})L^{\blq,\rm R}_{\substack{I\\J}}(\w)=if_{I}^{\blq}\d_{\substack{I\\J}}
-f_{I}^{\blq}\sum_{M}
K^{\blq}_{\substack{I\\ M}}
L^{\blq,\rm R}_{\substack{M\\J}}(\w).
\label{Lsuperbasisw2}
\ee
Since $f_{I}^{\blq}=0$ implies $L^{\blq,\rm 
R}_{\substack{I\\J}}=0$  we can solve Eq.~(\ref{Lsuperbasisw2}) in the 
subspace $\callS^{\blq}$ of superindices $I$ such that 
$f_{I}^{\blq}\neq 0$, and restrict the sum over $M$ to this subspace.
Notice that if $I\in \callS^{\blq}$ and $J\notin \callS^{\blq}$ then
$\d_{\substack{I\\J}}=0$ and therefore
Eq.~(\ref{Lsuperbasisw2}) becomes a homogeneous system of equations. 
Consequently, $L^{\blq,\rm 
R}_{\substack{I\\J}}$ is nonvanishing only for $I,J\in\callS^{\blq}$.
Let us split the superindices into two classes, 
one class with  $f_{I}^{\blq}>0$ and the other class with 
$f_{I}^{\blq}<0$. We order all vectors and matrices in such a way 
that the first entries correspond to superindices in the first class. 
Defining the matrices $\tilde{L}^{\blq}$ and $\tilde{K}^{\blq}$ 
according to~\cite{SRFHB.2011}
\be
L^{\blq,\rm R}_{\substack{I\\J}}
\equiv\sqrt{|f_{I}^{\blq}|}\,\tilde{L}^{\blq}_{\substack{I\\J}}
\sqrt{|f_{J}^{\blq}|}
\quad;\quad
\tilde{K}^{\blq}_{\substack{I\\J}}\equiv\sqrt{|f_{I}^{\blq}|}\,
K^{\blq}_{\substack{I\\J}}\sqrt{|f_{J}^{\blq}|},
\label{deflk}
\ee
we can rewrite Eq.~(\ref{Lsuperbasisw2}) as follows
\be
\left[(\w-\w^{\blq})\s_{z}^{\blq}+\tilde{K}^{\blq}\right]
\tilde{L}^{\blq}=i\mathbbm{1},
\label{casida2}
\ee
where $\mathbbm{1}$ is the identity matrix, $\w^{\blq}$ is the diagonal 
matrix with entries $\w^{\blq}_{I}$ and
\be
(\s_{z}^{\blq})_{\substack{I\\ J}}={\rm 
sign}(f^{\blq}_{I})\d_{\substack{I\\ J}}\;.
\nn
\ee
Since $\tilde{K}^{\blq}$ is hermitian we see from 
Eq.~(\ref{casida2}) that  $\tilde{L}^{\blq}$ is anti-hermitian, i.e.,
$\tilde{L}^{\blq}_{\substack{I\\ 
J}}=-\tilde{L}^{\blq\ast}_{\substack{J\\ I}}$, as it should.
Let us denote by $\W^{\l\blq}$ the values of $\w$ for which the 
matrix in the square brackets of Eq.~(\ref{casida2}) is singular and by 
$\tilde{Y}^{\l\blq}$ the vector belonging to the null space of the 
singular matrix:
\be
(\s_{z}^{\blq}\w^{\blq}-\tilde{K}^{\blq})\tilde{Y}^{\l\blq}
=\W^{\l\blq}\s_{z}^{\blq}\tilde{Y}^{\l\blq}.
\label{pseudoeigen}
\ee
For systems in equilibrium $\w^{\blq}_{I}\lessgtr 0$ implies that
$f^{\blq}_{I}\gtrless 0$. This property guarantees that the
$\W^{\l\blq}$'s are all real and can be arranged in pairs with entries 
of opposite sign. The 
reality of the $\W^{\l\blq}$'s is no longer guaranteed in stationary excited 
states (or in admixtures of them). However, if the pump is weak, as 
it is the case of 2PPE experiments,\cite{SBN.2004,Weinelt-book,UG.2007} 
then the {\em qp} occupations differ from their 
equilibrium values by a small amount and the $\W^{\l\blq}$'s continue to be real 
(although they cannot be arranged in pairs any longer).  
Under the assumption of reality we can normalize the $\tilde{Y}$ 
vectors according to
\be
\tilde{Y}^{\l\blq\ast}_{I}(\s_{z}^{\blq})_{\substack{I\\ J}}
\tilde{Y}^{\l'\blq}_{J}=
[\tilde{Y}^{\l\blq}]^{\dag}\s_{z}^{\blq}\tilde{Y}^{\l'\blq}=
s_{\l}\d_{\l\l'},
\label{normalization}
\ee
where $s_{\l}$ can be either $1$ or $-1$.
From Eq.~(\ref{pseudoeigen}) and from the normalization condition in 
Eq.~(\ref{normalization}) it is easy to show that the solution of 
Eq.~(\ref{Lsuperbasisw2}) with $I,J\in\callS^{\blq}$ can be written as
\be
L^{\blq,\rm R}_{\substack{I\\J}}(\w)=i
\sum_{\l}Y^{\l\blq}_{I}\frac{s_{\l}}{\w-\W^{\l\blq}+i\h}Y^{\l\blq\ast}_{J},
\label{spectralLR}
\ee
where $Y^{\l\blq}_{I}\equiv \sqrt{|f_{I}^{\blq}|}\;
\sum_{\l}\tilde{Y}^{\l\blq}_{I}$. The advanced component can be 
obtained similarly and differs from Eq.~(\ref{spectralLR}) only for 
the sign of the infinitesimal imaginary part of the denominator. 
Notice that the matrices $L^{\blq,\rm R/A}$ are manifestly anti-hermitian for real 
$\w\pm i\eta$, as it should. It is also easy to verify that 
in the noninteracting case 
Eq.~(\ref{spectralLR}) reduces to 
$\d_{\substack{I\\J}}\ell^{\blq,\rm R/A}_{I}$ [see Eq.~(\ref{xrasuperindex})].

\subsubsection{Lesser and Greater component}

Let us define the diagonal matrix 
$\ell_{\substack{I\\J}}=\d_{\substack{I\\J}}\ell_{I}$. Extracting the 
greater/lesser component of Eq.~(\ref{Lsuperbasis}) and Fourier 
transforming one finds (omitting the dependence on 
frequency)
\be
\left[\mathbbm{1}-i\ell^{\blq,\rm R}K^{\blq}\right]L^{\blq,\lessgtr}
=\ell^{\blq,\lessgtr}\left[\mathbbm{1}+iK^{\blq}L^{\blq,\rm A}\right].
\label{L<>}
\ee
We emphasize that this is an equation in the full space 
of superindices, i.e., matrix multiplication involves also 
superindices not belonging to $\callS^{\blq}$.
With the help of Eq.~(\ref{Lsuperbasisw2})
we can solve for $L^{\blq,\lessgtr}$ and
find
\be
L^{\blq,\lessgtr}=
(\mathbbm{1}+iL^{\blq,\rm R}K^{\blq})
\ell^{\blq,\lessgtr}(\mathbbm{1}+iK^{\blq}L^{\blq,\rm A}).
\nn
\ee
At difference with the retarded/advanced components, the 
lesser/greater components are nonvanishing also for 
indices $I,J\notin\callS^{\blq}$. For instance, the lesser 
two-particle correlator is given by
\bea
L^{\blq,<}_{\substack{I\\J}}\!\!&=&\!\!\d_{\substack{I\\J}} 
\ell^{\blq,<}_{I}
,\quad\quad\quad\quad\quad\quad\quad\quad\quad\quad\quad\quad\;\;\, I,J\notin\callS^{\blq}
\nn \\
L^{\blq,<}_{\substack{I\\J}}\!\!&=&\!\!i \ell^{\blq,<}_{I}
(K^{\blq}
L^{\blq,\rm A})_{\substack{I\\J}}
,\quad\quad\quad\quad\quad\quad\, I \notin\callS^{\blq},J\in\callS^{\blq}
\nn \\
L^{\blq,<}_{\substack{I\\J}}\!\!&=&\!\!i
(L^{\blq,\rm R}
K^{\blq})_{\substack{I\\J}}
\ell^{\blq,<}_{J}
,\quad\quad\quad\quad\quad\quad\, I \in\callS^{\blq},J\notin\callS^{\blq}
\nn \\
L^{\blq,<}_{\substack{I\\J}}\!\!&=&\!\!-\sum_{M\notin\callS^{\blq}}
(L^{\blq,\rm R}
K^{\blq})_{\substack{I\\M}}
\ell^{\blq,<}_{M}(K^{\blq}L^{\blq,\rm A})_{\substack{M\\J}}
\nn\\
\!\!&+&\!\!2\eta
\sum_{\a\b\blp\in\callS^{\blq}} 
L^{\blq,\rm R}_{\substack{I\\\b\a\blp}}
\frac{f_{\a\blp+\blq}\bar{f}_{\b\blp}}{(f^{\blq}_{\a\b\blp})^{2}}
L^{\blq,\rm A}_{\substack{\a\b\blp\\J}},
\quad\;\;\;\, I,J\in\callS^{\blq}
\nn\\
\label{Lexact}
\eea
where in the second term of the last equality we used
\be
\ell^{\blq,<}_{\a\b\blp}=2\eta\;
\ell^{\blq,\rm R}_{\a\b\blp}\;
\frac{f_{\a\blp+\blq}\bar{f}_{\b\blp}}{(f^{\blq}_{\a\b\blp})^{2}}
\;\ell^{\blq,\rm A}_{\a\b\blp},
\label{x<dec}
\ee
as it follows from the explicit expressions in Eqs.~(\ref{x<}) 
and (\ref{xrasuperindex}) and from the identity 
$\h/(\w^{2}+\h^{2})=\p\d(\w)$.

Although every term can be explicitly calculated we here make an 
approximation that is well justified in the physical regime we are 
working, i.e., the regime of weak pumps. In this regime the {\em qp} occupations 
$f_{\m\blk}$ are either close to zero or close to 1. If 
$I=(\m\n\blk)\notin\callS^{\blq}$ then [see Eq.~(\ref{superocc})]
$f^{\blq}_{I}=f_{\n\blk}-f_{\m\blk+\blq}=0$ which implies that both
$f_{\n\blk}$ and $f_{\m\blk+\blq}$ are either close to zero or 
close to 1 and hence that both products 
$f_{\n\blk}\bar{f}_{\m\blk+\blq}$  and 
$\bar{f}_{\n\blk}f_{\m\blk+\blq}$ are close to zero. Taking into 
account Eqs.~(\ref{x>}) and (\ref{x<}) we then see 
that  $\ell^{\blq,\lessgtr}_{I}$ is small for $I\notin\callS^{\blq}$. 
Approximating  
\be
\ell^{\blq,\lessgtr}_{I}\simeq 0\quad\quad{\rm 
for}\;I\notin\callS^{\blq},
\nn
\ee
we can write for all $I$ and $J$ 
\be
L^{\blq,<}_{\substack{I\\J}}(\w)=-2\eta
\sum_{\a\b\blp\in\callS^{\blq}} 
L^{\blq,\rm R}_{\substack{I\\\b\a\blp}}(\w)
\frac{f_{\a\blp+\blq}\bar{f}_{\b\blp}}{(f^{\blq}_{\a\b\blp})^{2}}
L^{\blq,\rm A}_{\substack{\a\b\blp\\J}}(\w).
\label{L<>approx2}
\ee
We now insert in Eq.~(\ref{L<>approx2}) the spectral decomposition 
for the retarded/advanced two-particle correlator, see 
Eq.~(\ref{spectralLR}).
The resulting double sum over $\l,\l'$ can be split into a sum over $\l=\l'$ and a sum over 
$\l\neq\l'$. In the limit $\eta\to 0$ the latter 
is finite whereas the former yields a sum of $\d$-functions.
We can then restrict the sum to $\l=\l'$ and get
\be
L^{\blq,<}_{\substack{I\\J}}(\w)=2\p\,\sum_{\l}
F^{\l\blq}\,Y^{\l\blq}_{I}\,
\d(\w-\W^{\l\blq})\,
Y^{\l\blq\ast}_{J}\;\;,
\label{L<approx4}
\ee
where we have defined 
\be
F^{\l\blq}\equiv \sum_{\a\b\blp\in\callS^{\blq}} 
Y^{\l\blq\ast}_{\a\b\blp}\,
\frac{f_{\a\blp+\blq}\bar{f}_{\b\blp}}{(f^{\blq}_{\a\b\blp})^{2}}
\,Y^{\l\blq}_{\a\b\blp},
\nn
\ee
and introduced the convention $Y^{\l\blq}_{I}=0$ for $I\notin 
\callS^{\blq}$.
A similar expression can be derived for the greater component
\be
L^{\blq,>}_{\substack{I\\J}}(\w)=2\p\,\sum_{\l}
\bar{F}^{\l\blq}\,Y^{\l\blq}_{I}\,
\d(\w-\W^{\l\blq})\,
Y^{\l\blq\ast}_{J}\;\;,
\label{L>approx4}
\ee
where we have defined 
\be
\bar{F}^{\l\blq}\equiv \sum_{\a\b\blp\in\callS^{\blq}} 
Y^{\l\blq\ast}_{\a\b\blp}\,
\frac{\bar{f}_{\a\blp+\blq}f_{\b\blp}}{(f^{\blq}_{\a\b\blp})^{2}}
\,Y^{\l\blq}_{\a\b\blp}.
\nn
\ee
We have verified that Eqs.~(\ref{L<approx4}) and (\ref{L>approx4}) 
reduce to $\ell^{\blq,\lessgtr}(\w)$ in the noninteracting case and that  in 
equilibrium we recover the fluctuation dissipation theorem.

\subsection{Excited self-energy and Green's function}
\label{Sigma-G-sec}

Let us evaluate 
$\S_{\blx_{1}\blx_{4}}(z,z')$ in 
Fig.~\ref{sel}~(a). 
Expanding the self-energy analogously to the Green's function [see 
Eq.~(\ref{gexp})], i.e.,
\be
\S_{\blx_{1}\blx_{4}}(z,z')=
\sum_{pq}\vf_{p}(\blx_{1})\vf_{q}^{\ast}(\blx_{4})\S_{pq}(z,z'),
\nn
\ee
and taking into account the expansion of $L$ in Eq.~(\ref{Lexp}) as 
well as the definition of the four-index screened interaction  in 
Eq.~(\ref{4indx}), it is a matter of simple algebra to find
\be
\S_{pq}(z,z')=-i^{2}\sum_{ijmnk}g_{k}(z,z')\,W_{\substack{pk\\ji}}\,
L_{\substack{ij\\mn}}(z,z')W_{\substack{nm\\kq}},
\nn
\ee
where $W_{\substack{pk\\ji}}\equiv W_{pjki}$
(in analogy with the definition of the kernel $K$ in Eq.~(\ref{Lgeneralbasis})).
Extracting the lesser/greater 
component, Fourier transforming and using Eqs.~(\ref{g0}) we find
\begin{subequations}
\begin{alignat}{2}    
&\S_{pq}^{<}(\w)=i
\sum_{ijmnk}f_{k}\,W_{\substack{pk\\ji}}\,
L_{\substack{ij\\mn}}^{<}(\w-\e_{k})W_{\substack{nm\\kq}},
\nn \\
&\S_{pq}^{>}(\w)=-i
\sum_{ijmnk}\bar{f}_{k}\,W_{\substack{pk\\ji}}\,
L_{\substack{ij\\mn}}^{>}(\w-\e_{k})W_{\substack{nm\\kq}}.
\nn
\end{alignat}   
\end{subequations}
We make explicit the dependence on the band-spin indices and 
momenta. Due to momentum conservation 
$\S_{\m\blp\n\blp'}=\d_{\blp\blp'}\S_{\m\n\blp}$. After some algebra 
the lesser self-energy takes the form
\be
\S^{<}_{\m\n\blp}(\w)=i\!\!\sum_{IJ,\g\blq}\!f_{\g\blp-\blq}W^{\blq}_{\subalign{&\m\g\blp-\blq\\&I}}
L^{\blq,<}_{\subalign{&I\\&J}}(\w-\e_{\g\blp-\blq})
W^{\blq}_{\subalign{&J\\&\g\n\blp-\blq}}
\label{se<1}
\ee
with a similar expression for the greater self-energy.
In Eq.~(\ref{se<1}) the sum is restricted to $I,J\in \callS^{\blq}$ 
due to the approximation 
in Eq.~(\ref{L<>approx2}), according to which $L^{\blq,<}_{\substack 
{I\\J}}$ vanishes if $I$ and/or $J$ do not belong to $\callS^{\blq}$. 
Inserting the expansion in Eq.~(\ref{L<approx4}) we get
\bea
\S^{<}_{\m\n\blp}(\w)\!\!&=&\!\!2\p i\sum_{\l}
\sum_{IJ,\g\blq}f_{\g\blp-\blq} F^{\l\blq}W^{\blq}_{\subalign{&\m\g\blp-\blq\\&I}}\,
Y^{\l\blq}_{I}
\nn\\
\!\!&\times&\!\!
\d(\w-\e_{\g\blp-\blq}-\W^{\l\blq})\,
Y^{\l\blq\ast}_{J}W^{\blq}_{\subalign{&J\\&\g\n\blp-\blq}}.
\nn
\eea
Following similar steps the greater  
self-energy reads
\bea
\S^{>}_{\m\n\blp}(\w)\!\!&=&\!\!-2\p i\sum_{\l}
\sum_{IJ,\g\blq}\bar{f}_{\g\blp-\blq}\bar{F}^{\l\blq}\,W^{\blq}_{\subalign{&\m\g\blp-\blq\\&I}}\,
Y^{\l\blq}_{I}
\nn\\
\!\!&\times&\!\!
\d(\w-\e_{\g\blp-\blq}-\W^{\l\blq})\,
Y^{\l\blq\ast}_{J}W^{\blq}_{\subalign{&J\\&\g\n\blp-\blq}},
\nn
\eea
and hence the 
retarded/advanced self-energy follows from the Hilbert transform
\bea
\S^{\rm R/A}_{\m\n\blp}(\w)
&=&
\sum_{\l}\sum_{IJ,\g\blq}
W^{\blq}_{\subalign{&\m\g\blp-\blq\\&I}}\,
Y^{\l\blq}_{I}
\nn\\
&\times&
\frac{\bar{f}_{\g\blp-\blq}\bar{F}^{\l\blq}+f_{\g\blp-\blq}F^{\l\blq}}
{\w-\e_{\g\blp-\blq}-\W^{\l\blq}\pm i\h}\,Y^{\l\blq\ast}_{J}
W^{\blq}_{\subalign{&J\\&\g\n\blp-\blq}}.\quad\;\;
\label{mainsigma}
\eea
Equation~(\ref{mainsigma}) does not contain any empirical parameter;  it 
provides the nonequilibrium self-energy in terms of quantities that 
can all be obtained {\em ab initio}.

As the self-energy is diagonal in  
momentum space  the dressed Green's function $G$ is 
diagonal too. Therefore,  it is convenient to manipulate matrices 
with indices only in the band-spin sector. We define
$(\S_{\blk})_{\a\b}\equiv\S_{\a\b\blk}$, 
$(\e_{\blk})_{\a\b}\equiv\d_{\a\b}\e_{\a\blk}$, and 
$(G_{\blk})_{\a\b}\equiv G_{\a\b\blk}$.
Then, the retarded Green's function can be calculated from
\be
G^{\rm R/A}_{\blk}(\w)=\frac{1}{\w-\e_{\blk}-\S^{\rm R/A}_{\blk}(\w)}.
\label{graexact}
\ee
Experiments\cite{CRCZGBBKGP.2012,SYACFKS.2012,NOHFMEAABEC.2014,BVLNL.2014} 
and numerical 
simulations~\cite{sm.2015,SDCMCM.2016} indicate that the electron 
occupations in the quasi-stationary 
excited state follow a 
Fermi-Dirac distribution with temperatures $T_{\a}$ and chemical 
potentials $\m_{\a}$ depending on the band-spin index $\a$. 
Of course $T_{\a}$ and  $\m_{\a}$ vary on a picosecond time-scale but 
they can be considered as constant on the time-scale of the probe 
pulse. From this evidence we infer  that the 
recombination of electrons with different band-spin index $\a$ is 
severely suppressed and that the lesser Green's function fulfills the 
approximate fluctuation-dissipation relation
\be
G^{<}_{\a\b\blk}(\w)=-\d_{\a\b}f_{\a}(\w)[G^{\rm R}_{\a\a\blk}(\w)-G^{\rm 
A}_{\a\a\blk}(\w)],
\label{heuristic6}
\ee
where $f_{\a}(\w)=1/(e^{(\w-\m_{\a})/T_{\a}}+1)$. The $\a$-dependent 
temperature and chemical potential can be extracted by a best 
fitting of the electronic populations as obtained from, e.g., the 
one-time Kadanoff-Baym propagation.\cite{sm.2015} Using the Green's 
function of Eq.~(\ref{heuristic6}) in Eq.~(\ref{photocurr2}) the photocurrent follows.

This concludes our first-principle diagrammatic approach to deal with 
excitonic features in TR-PE spectra. 
In the next Section we study excitonic features in a minimal 
model  and assess the 
accuracy of the proposed theory.

\section{Application to a Minimal Model}

We consider a 
one-dimensional insulator of length $\callL$ with 
one valence 
band and one conduction band separated by a direct gap of strength 
 $\D$.\cite{YLU.2012} Since the formation of excitons is due to the attraction 
between a valence hole and a conduction electron we discard the 
Coulomb interaction between electrons in the same band. For 
simplicity we also discard spin. Thus, the Hamiltonian of the 
insulator reads
\bea
\hat{H}_{\rm ins}&=&\sum_{k}(\e_{vk}\hat{v}^{\dag}_{k}\hat{v}_{k}
+\e_{ck}\hat{c}^{\dag}_{k}\hat{c}_{k})-
U(0)\frac{N_{v}}{\callL}\sum_{k}\hat{c}^{\dag}_{k}\hat{c}_{k}
\nn\\
&+&
\frac{1}{\callL}\sum_{k_{1}k_{2}q}U(q)\,\hat{v}^{\dag}_{k_{1}+q}\hat{c}^{\dag}_{k_{2}-q}
\hat{c}_{k_{2}}\hat{v}_{k_{1}},
\label{minmodham}
\eea
where $\hat{v}_{k}$ ($\hat{c}_{k}$) annihilates an electron of momentum $k$ in the 
valence (conduction) band and $W(q)\equiv U(q)/\callL$ is the  statically screened 
interaction. The last term in the first row represents the interaction 
of a conduction electron with the positive background in the valence 
band, $N_{v}$ being the number of protons (which is also equal to the 
number of valence electrons in the ground state). For this model
the ground state is obtained by filling all single-particle valence 
states with one electron. Hence the 
interaction between the valence background and the conduction 
electrons vanishes.

\subsection{Analytic treatment for a single bound exciton}
\label{analytic-solution-sec}

\label{lessec}

The insulator Hamiltonian commutes with the total 
number of conduction electrons 
$\hat{N}_{c}=\sum_{k}\hat{c}^{\dag}_{k}\hat{c}_{k}$ and with the 
total number of valence electrons 
$\hat{N}_{v}=\sum_{k}\hat{v}^{\dag}_{k}\hat{v}_{k}$.
We consider the special case of a stationary excited state of 
vanishing total momentum  with   
one electron in the conduction band (and hence with one hole in the 
valence band). Denoting by 
$|\Q_{g}\ket=\prod_{k}\hat{v}^{\dag}_{k}|0\ket$
the ground state of energy $E_{g}$ we write this excited state as
\be
|\Q\ket=\sum_{k}Y_{k}\hat{c}^{\dag}_{k}\hat{v}_{k}|\Q_{g}\ket=
\sum_{k}Y_{k}|\F_{k}\ket,
\label{exexp}
\ee
where we introduced the {\em eh} states 
$|\F_{k}\ket\equiv\hat{c}^{\dag}_{k}\hat{v}_{k}|\Q_{g}\ket$.
It is a matter of straightforward algebra to show that 
$\hat{H}_{\rm ins}|\Q\ket$ is again a linear combination of the 
$|\F_{k}\ket$'s. The possible excited state energies 
$E=E_{g}+\W$ are found by solving the eigenvalue problem 
\be
(\w_{k-q}-\W)Y_{k}=\frac{1}{\callL}\sum_{q}U(q)Y_{k-q},
\label{bcs}
\ee
with $\w_{k}\equiv \e_{ck}-\e_{vk}>\D= \e_{c0}-\e_{v0}$. 
For a momentum independent interaction $U(q)=U>0$ the expansion 
coefficients have the form
\be
Y_{k}=\frac{\sqrt{R}}{\w_{k}-\W},
\label{ansatz}
\ee
where the positive constant $R$ is fixed by the normalization 
$\sum_{k}|Y_{k}|^{2}=1$. Equation~(\ref{bcs}) has a continuum of 
solutions $\W=\D-b$ with $b<0$ and
one split-off solution $\W_{X}=\D-b_{X}$ with binding energy $b_{X}>0$. The latter corresponds to a bound 
{\em eh} state or exciton.
Notice that for any arbitrary small but 
finite $U$ the excitonic amplitude $Y_{k}\sim 1/\sqrt{\callL}$
for $\callL\to\iif$
whereas $b_{X}$ converges to a finite 
positive value.

By definition, the lesser Green's function of the system in the 
exciton state $|\Q\ket=|\Q_{X}\ket$ is 
\bea
G_{cc,k}^{<}(t,t')\!\!&=&\!\!i\bra\Q_{X}|\hat{c}^{\dag}_{kH}(t')\hat{c}_{kH}(t)|\Q_{X}\ket
\nn\\
\!\!&=&\!\!
i\bra\Q_{X}|\hat{c}^{\dag}_{k}e^{-i(\hat{H}-E_{g}-\W_{X})(t'-t)}
\hat{c}_{k}|\Q_{X}\ket.
\nn
\eea
The only many-body states having a nonvanishing 
overlap with $\hat{c}_{k}|\Q_{X}\ket$ are the states 
$\hat{v}_{p}|\Q_{g}\ket$ which are also eigenstates 
$\hat{H}_{\rm ins}$ with eigenvalue $E_{g}-\e_{vp}$. Inserting a 
completeness relation to the right of $\hat{c}^{\dag}_{k}$ and Fourier 
transforming we find the {\em exact} result
\be
G_{cc,k}^{<}(\w)=2\p i|Y_{k}|^{2}\d(\w-\W_{X}+\e_{vk}).
\label{FexG<cc0}
\ee
In the following we show that our diagrammatic approach yields 
precisely Eq.~(\ref{FexG<cc0}). Before, however, we observe that 
substitution of Eq.~(\ref{FexG<cc0}) into 
Eq.~(\ref{photocurr3}) leads to the photocurrent
\be
I(\blk)=2\p|Y_{k}a_{0}D_{\blk}|^{2}
\d(\w_{0}+\W_{X}+\e_{vk}-\e_{f\blk}),
\label{pcmodelex}
\ee
where, without any loss of generality, we took $\w_{0}>0$  (in this 
case $G_{cc,k}^{<}(\e_{f\blk}+\w_{0})$ does not contribute). 
Equation~(\ref{pcmodelex}) 
agrees with Eq.~(\ref{heupc}), as it should.

To calculate the (dressed) excited lesser Green's function diagrammatically we 
need an excited {\em qp} Green's function $g$. Here we evaluate $g$ 
in the 
HF approximation. The excited noninteracting Green's function 
$g^{(0)}$
with one conduction electron and one valence hole in the lowest 
energy state reads
\begin{subequations}
\begin{alignat}{2}
&g^{(0),<}_{vv,k}(\w) = 2\p i\,\bar{\d}_{k0} \,\d(\w-\e_{vk}),
\\
&g^{(0),>}_{vv,k}(\w) = -2\p i\,\d_{k0} \,\d(\w-\e_{vk}),
\\
&g^{(0),<}_{cc,k}(\w) = 2\p i\,\d_{k0} \,\d(\w-\e_{ck}+U(0)N_{v}/\callL),
\\
&g^{(0),>}_{cc,k}(\w) = -2\p i\,\bar{\d}_{k0} \,\d(\w-\e_{ck}+U(0)N_{v}/\callL),
\end{alignat}
\label{gnoint}
\end{subequations}
\vspace{-0.2cm}

\noindent
and $g^{(0),\lessgtr}_{cv,k}=g^{(0),\lessgtr}_{vc,k}=0$. In Eqs.~(\ref{gnoint})
we defined $\bar{\d}_{k0}=1-\d_{k0}$. The HF potential 
contains only the Hartree part since the interaction preserves the 
band-spin index and $g^{(0)}$ is diagonal. Using Eqs.~(\ref{gnoint}) 
one finds 
\be
V_{{\rm HF},\a\a k}=\d_{\a c}\frac{U(0)}{\callL}N_{v}
+\callO(1/\callL).
\label{HFpot}
\ee
Accordingly, the excited HF Green's function is
\begin{subequations}
\begin{alignat}{2}
&g^{<}_{vv,k}(\w)=2\p i\,\bar{\d}_{k0} \,\d(\w-\e_{vk}),
\label{HFg<vv}\\
&g^{>}_{vv,k}(\w)=-2\p i\,\d_{k0} \,\d(\w-\e_{v0}),
\\
&g^{<}_{cc,k}(\w)=2\p i\,\d_{k0} \,\d(\w-\e_{c0}),
\label{mathfrackg<}
\\
&g^{>}_{cc,k}(\w)=-2\p i\,\bar{\d}_{k0} \,\d(\w-\e_{ck}).
\label{HFg>cc}
\end{alignat}
\label{HFg}
\end{subequations}
\vspace{-0.4cm}

\noindent
We observe that if we used 
the HF $g^{<}_{cc,k}$ to evaluate the 
photocurrent in Eq.~(\ref{photocurr3}) we would find
\be
I(\blk)=2\p|a_{0}D_{\blk}|^{2}\d_{k0}\d(\w_{0}+\e_{c0}-\e_{f\blk}),
\nn
\ee
which coincides with the noninteracting limit of 
Eq.~(\ref{pcmodelex}), i.e., $Y_{k}=\d_{k0}$ and $b_{X}=0$. 
As expected the HF approximation (and any other {\em qp} approximation) 
does not capture the exciton peak in the
energy-resolved and angle-resolved photocurrent.
\begin{figure}[t]  
\includegraphics*[width=.46\textwidth]{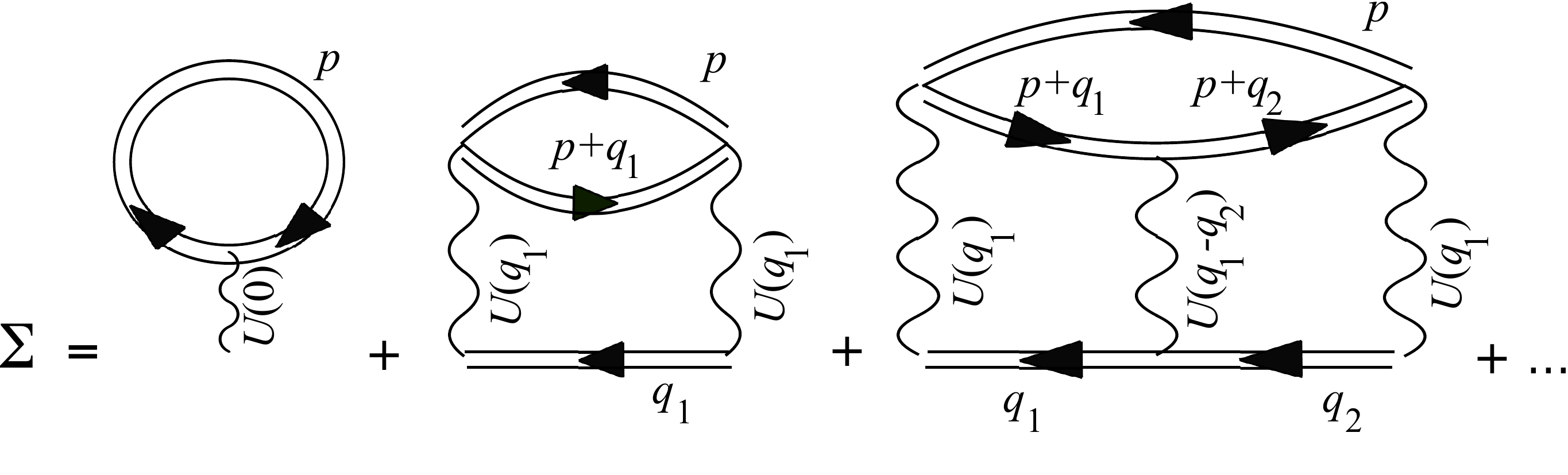}  
\caption{Self-energy diagrams for the model Hamiltonian of Eq.~(\ref{minmodham}).}  
\label{diagramsfig}  
\end{figure}

For the model Hamiltonian in Eq.~(\ref{minmodham}) the self-energy diagrams 
of Fig.~\ref{sel}~(a) that contain a polarization insertion
vanish. Thus, 
we only need to evaluate the
self-energy diagrams in Fig.~\ref{diagramsfig}, with the exception of 
the first (Hartree) 
diagram. Since we are interested in $G_{cc}$ 
and since the self-energy has vanishing $cv$ and $vc$ components we 
only calculate the $cc$ component. For simplicity we also consider 
the case  of
vanishing momentum $k=0$ and a momentum independent interaction $U(q)=U$. 
We have
$\S_{cc,0}(z,z')\equiv \S(z,z')-\S_{\rm H}(z,z')$ where 
$\S$ is the full series of Fig.~\ref{diagramsfig} and $\S_{\rm 
H}$  is the first diagram of the series. Introducing the 
averaged {\em eh} propagator
\be
\ell_{p}(z,z')=\frac{1}{\callL}\sum_{q}g_{cc,q}(z,z')g_{vv,p+q}(z',z),
\label{exprop}
\ee
we can write the full series as
\be
\S(z,z')=-\frac{i}{\callL}\sum_{p}T_{p}(z,z')g_{vv,p}(z,z').
\label{kelsigma}
\ee
where we have defined the $T$-matrix
\be
T_{p}(z,z')\equiv U\d(z,z')+iU\int dz_{1}\,\ell_{p}(z,z_{1})T_{p}(z_{1},z').
\label{tmatk}
\ee

To calculate the lesser and greater components of $\S$ (which are necessary to 
calculate $G^{<}_{cc,0}$) we need the 
lesser and greater components of $T_{p}$. This can be achieved 
without going through the spectral decomposition of Section~\ref{BSELsec} since the 
system is in a pure (excited) state which
is simple enough. The spectral decomposition will be used in the 
next Section where we consider the system in an  
admixtures of excited states. Using the Langreth rules 
in Eq.~(\ref{tmatk}) we get
\be
T^{\lessgtr}_{p}(\w)=i\frac{U^{2}}{|1-iU\ell^{\rm 
R}_{p}(\w)|^{2}}\ell^{\lessgtr}_{p}(\w).
\label{tmat<>2}
\ee
From the definition of the {\em eh} propagator in Eq.~(\ref{exprop}) 
and using the excited HF Green's functions  in Eqs.~(\ref{HFg})
we find 
$\ell^{<}_{p}(\w)=(2\p/\callL)\d_{p0}\,\d(\w-\D)$.
Therefore $T^{<}_{p}(\w)\propto \d_{p0}$ and consequently the lesser self-energy 
\be
\S^{<}(t,t')=-\frac{i}{\callL}T^{<}_{0}(t,t')g^{<}_{vv,0}(t,t')=0.
\nn
\ee
Thus we only need to evaluate the
greater self-energy. From  
Eq.~(\ref{kelsigma}) 
\bea
\S^{>}(\w)&=&-\frac{i}{\callL}\sum_{p}\int\frac{d\w'}{2\p}T^{>}_{p}(\w-\w')
\mathfrak{g}^{>}_{vv,p}(\w')
\nn\\
&=&-\frac{1}{\callL}T^{>}_{0}(\w-\e_{v0}).
\label{se>mm}
\eea
It is important to emphasize that if we had used a {\em ground state} 
$g$ then also $\S^{>}=0$ since there would be no holes 
in the valence band and hence $g^{>}_{vv,p}=0$. 
The calculation of $T^{>}_{0}$ requires the explicit form of 
$\ell^{>}_{0}$ and $\ell^{\rm R}_{0}$. These follow from Eq.~(\ref{exprop}) 
\be
\ell^{>}_{0}(\w)=
\frac{2\p}{\callL}\sum_{q}\d(\w-\w_{q})
+\callO(1/\callL),
\label{x0>}
\ee
and
\be
\ell_{0}^{\rm R}(\w)=\frac{1}{\callL}
\sum_{q}\frac{i}{\w-\w_{q}+i\eta}
+\callO(1/\callL).
\label{x0R}
\ee
Substitution of these results into Eq.~(\ref{tmat<>2}) yields
\be
T^{>}_{0}(\w)=2i U \frac{y(\w)}{(1-x(\w))^{2}+y^{2}(\w)},
\nn
\ee
where we have defined $x(\w)\equiv\Re[iU\ell_{0}^{\rm R}(\w)]$ and 
$y(\w)\equiv\Im[iU\ell_{0}^{\rm R}(\w)]=(U/2)\ell^{>}_{0}(\w)$.
The quantity $y(\w)$ vanishes for 
$\w<\D$, see Eq.~(\ref{x0>}). However, this {\em does not imply}
that $T_{0}^{>}(\w)$ vanishes in the same region. In fact,
\be
\lim_{y\ra 0^{+}}
\frac{y}{(1-x)^{2}+y^{2}}=\p\d(1-x),
\nn
\ee
and hence $T_{0}^{>}(\w)$ is nonvanishing for $\w<\D$ if  
in this frequency region $1-x(\w)=0$.  From Eq.~(\ref{x0R}) we have
\be
1-x(\w)=1+\frac{U}{\callL}\sum_{q}\frac{1}{\w-\w_{q}}=0.
\nn
\ee
This equation is identical to Eq.~(\ref{bcs}) after the renaming 
$\w=\W$. Thus $1-x(\w)=0$ 
has a continuum of solutions for $\w>\D$ and one split-off 
solution at $\w=\W_{X}<\D$. Therefore $T^{>}_{0}(\w)$
can be conveniently rewritten as
\bea
T^{>}_{0}(\w)&=&
\frac{2\p i U}{\left|\de 
x(\w)/\de\w\right|_{\w=\W_{X}}}\d(\w-\W_{X})
\nn\\
&+&2i U\,{\rm Reg}
\left[\frac{y(\w)}{(1-x(\w))^{2}+y^{2}(\w)}\right],
\label{tmat>0final}
\eea
where  ${\rm Reg}$ denotes the nonsingular part of the function.

We can now evaluate $\S^{>}$ from Eq.~(\ref{se>mm}) as well as
the retarded self-energy
\be
\S^{\rm 
R}_{cc,0}(\w)=-\frac{i}{\callL}\int\frac{d\w'}{2\p}\frac{T^{>}_{0}(\w'-\e_{v0})}{\w-\w'+i\eta}.
\label{sigR0}
\ee
The Hartree part does not contribute to $\S^{>}$ and it 
is therefore correctly removed in Eq.~(\ref{sigR0}). 
Using Eq.~(\ref{tmat>0final})  we 
find
\be
\S^{\rm R}_{cc,0}(\w)=\frac{R_{X}}{\w-\W_{X}-\e_{v0}+i\eta}+
\S_{\rm reg}^{\rm R}(\w),
\label{sigmacc0R}
\ee
where
\be
R_{X}=\frac{U/\callL}{\left|\de 
x(\w)/\de\w\right|_{\w=\W_{X}}},
\nn
\ee
is the excitonic residue of the singular part whereas 
$\S_{\rm reg}^{\rm R}$ is the regular (nonsingular) part.
Both $R_{X}$ and $\S_{\rm reg}^{\rm R}$ scale like 
$1/\callL$ and are therefore infinitesimally small in the thermodynamic 
limit. Interestingly, $R_{X}$ is exactly the same constant that 
appears in the normalized excitonic amplitude of Eq.~(\ref{ansatz}).

From the retarded self-energy  the retarded Green's 
function follows
\be
G^{\rm R}_{cc,0}(\w)=\frac{1}{\w-\e_{c0}-\S^{\rm 
R}_{cc,0}(\w)}.
\nn
\ee
For $\w\simeq \e_{X}\equiv \W_{X}+\e_{v0}=\e_{c0}-b_{X}$ the self-energy is 
dominated by the first term in Eq.~(\ref{sigmacc0R}). 
Thus for frequencies in the neighborhood of  
$\e_{X}$ we can write
\bea
G^{\rm R}_{cc,0}(\w\sim \e_{X})&\simeq&
\frac{1}{\e_{X}-\e_{c0}-\frac{R_{X}}{\w-\e_{X}+i\eta}}
\nn\\
&=&\frac{R_{X}/b^{2}_{X}}{\w-\e_{X}+i\eta}+
\callO(1/\callL),
\nn
\eea
where we took into account that
$R_{X}\sim 1/\callL$. In the same neighborhood the spectral 
function $A=i[G^{\rm R}_{cc,0}-G^{\rm A}_{cc,0}]$ reads
\bea
A(\w\simeq \e_{X})
\simeq 2\p Z_{X}\,\d(\w-\e_{X}),
\nn
\eea
where we have defined the  
{\em excitonic qp weight} as
\be
Z_{X}\equiv \frac{R_{X}}{b_{X}^{2}}.
\nn
\ee
The physical meaning of $Z_{X}$ is the amount of  spectral weight 
that a bare excited electron transfers to the electron in the 
bound {\em eh} pair. We further observe that $Z_{X}$ 
is precisely the excitonic amplitude $|Y_{0}|^{2}$, see Eq.~(\ref{ansatz}).

To calculate the excited lesser Green's function we use 
Eq.~(\ref{heuristic6}), i.e., $G^{<}_{cc,0}(\w)=
if_{c}(\w)A(\w)$,
where $f_{c}(\w)$ is the Fermi function for the conduction band. 
To find the temperature $T_{c}$ and chemical potential $\m_{c}$  we 
observe that the occupations of the excited state are 
$f_{ck}=\d_{k0}$, see 
Eq.~(\ref{mathfrackg<}). Therefore $T_{c}=0$ and 
$\m_{c}$ is just above $\e_{c0}$.
From the 
previous analysis  we know that the spectral function has a 
$\d$-like peak in $\w=\e_{X}<\e_{c0}$ and it is otherwise smooth and 
nonvanishing for $\w>\e_{c0}$. More precisely the self-energy is 
responsible  for 
moving the noninteracting spectral peaks to the right by an amount 
$\simeq 1/\callL$. 
Therefore 
only the exciton peak is below $\m_{c}$ and the excited lesser Green's 
function reads
\be
G^{<}_{cc,0}(\w)=2\p iZ_{X}\d(\w-\e_{X}).
\nn
\ee
Since $Z_{X}=|Y_{0}|^{2}$ our diagrammatic approach yields the exact result of 
Eq.~(\ref{FexG<cc0}).

The analysis of this Section supports the validity of the proposed 
theoretical framework. In the next Section 
we consider stationary excited states with a smooth distribution of electrons 
in the conduction band and investigate the behavior of the exciton 
peak in different 
regimes.

\subsection{Numerical results at finite {\em eh} density}
\label{finite-t-sec}

In this Section we study the PE problem for
finite {\em eh} densities.
From Eq.~(\ref{minmodham}) and the definition in 
Eq.~(\ref{newKtensor}) with $K\to W$ we see that
\bea
W^{q}_{\subalign{&\m\n k\\& \a\b k'}}&=&
W^{q}_{\subalign{&\m k+q\,\n k\\& \a k'\,\b k'+q}}
\nn\\
&=&
W_{\m k+q,\a k',\n k,,\b k'+q}
\nn\\
&=&
\d_{\m\b}\d_{\a\n}[\d_{\m c}\d_{\a v}+\d_{\m v}\d_{\a c}]U/\callL.
\eea
Inserting this result into Eq.~(\ref{se<1}) and the analogous for the 
greater self-energy we obtain
\begin{subequations}
\begin{alignat}{2}    
&\S^{<}_{p}(\w)\equiv \S^{<}_{cc,p}(\w)=
iU^{2}\sum_{q}f_{v p-q} L^{q,<}(\w-\e_{v p-q}),
\nn\\
&\S^{>}_{p}(\w)\equiv \S^{>}_{cc,p}(\w)=
-iU^{2}\sum_{q}\bar{f}_{v p-q}L^{q,>}(\w-\e_{v p-q}),
\nn
\end{alignat}
\end{subequations}
where we defined 
\be
L^{q,\lessgtr}(\w)\equiv\frac{1}{\callL^{2}}\sum_{p_{1}p_{2}}
L^{q,\lessgtr}_{\subalign{&cvp_{1}\\&vcp_{2}}}(\w).
\label{Lq<tot}
\ee
In the calculations we solve Eq.~(\ref{pseudoeigen}) for different 
interaction strengths $U$ and occupations $f_{\a p}$.
We consider a valence band with energies in the interval $[-w/2,w/2]$ 
and dispersion $\e_{vk}=(w/2)\!\cos k$ and a conduction band with 
energies in the interval $[w/2+\D,3w/2+\D]$ and dispersion
$\e_{ck}=-(w/2)\!\cos k +w +\D$; $w>0$ is the bandwidth of both bands. 
The insulator has a direct gap of 
strength $\D$ at $k=0$.
The electron occupations $f_{\a k}$ in the excited state 
are Fermi-Dirac distributions
with the same temperature $T$
and {\it different} chemical potentials $\mu_{\a}$
\be
f_{\a k}=\frac{1}{e^{(\e_{\a k}-\m_{\a})/T}+1}, \quad \a=v,c \,.
\label{excpop}
\ee

\begin{figure}[tbp]  
\includegraphics*[width=.46\textwidth]{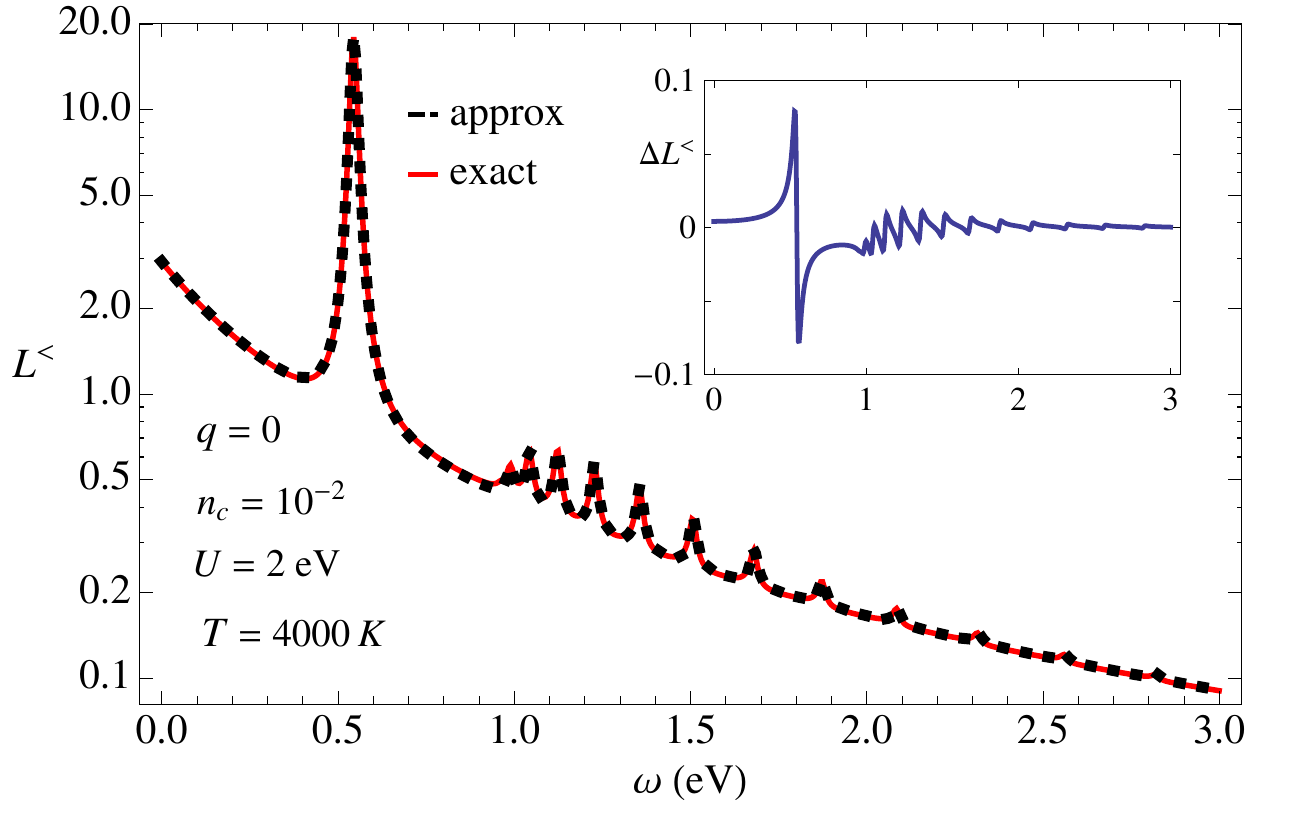}  
\caption{Log-plot of $L^{q,<}$ (in arbitrary units) at
$q=0$ according to  Eq. (\ref{L<approx4}) (dashed black line) 
and Eq. (\ref{Lexact}) (solid red line). The inset shows the 
difference between the two curves.}  
\label{Lless}  
\end{figure} 
\begin{figure}[bp]  
\includegraphics*[width=.31\textwidth]{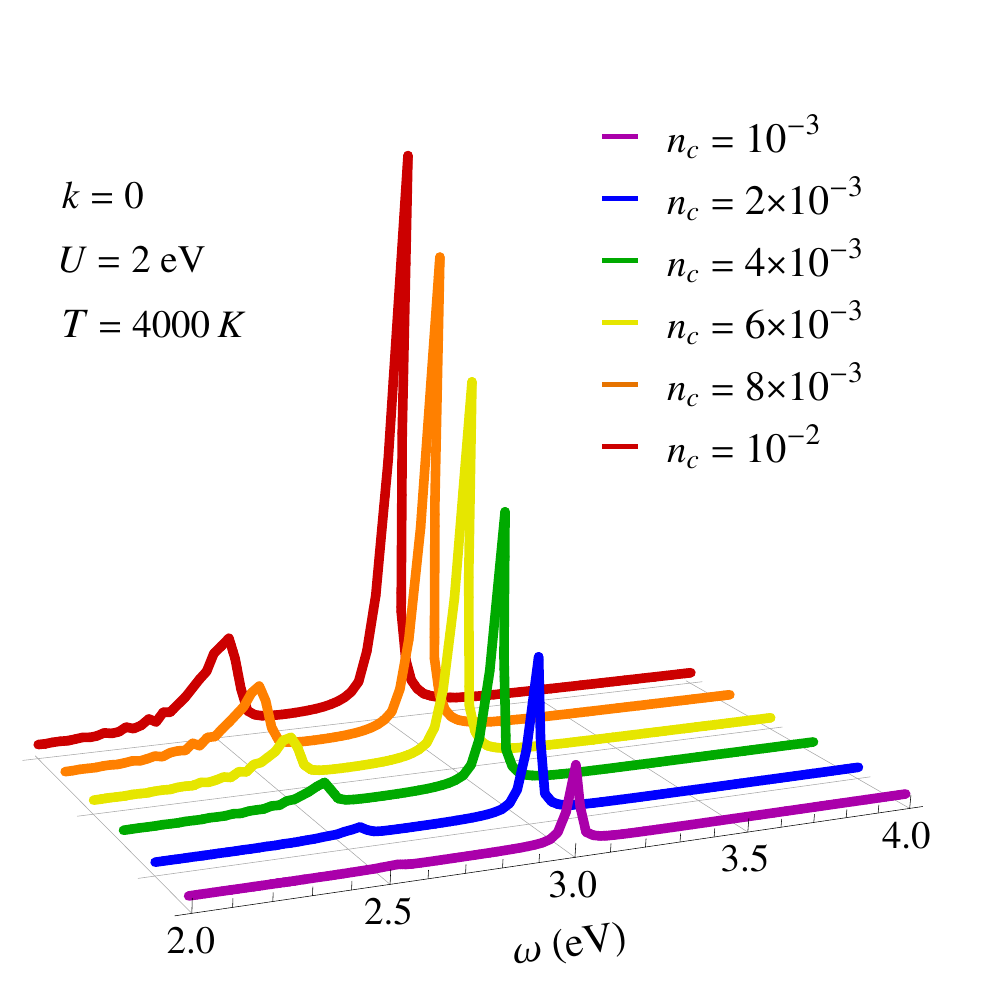}  
\includegraphics*[width=.15\textwidth]{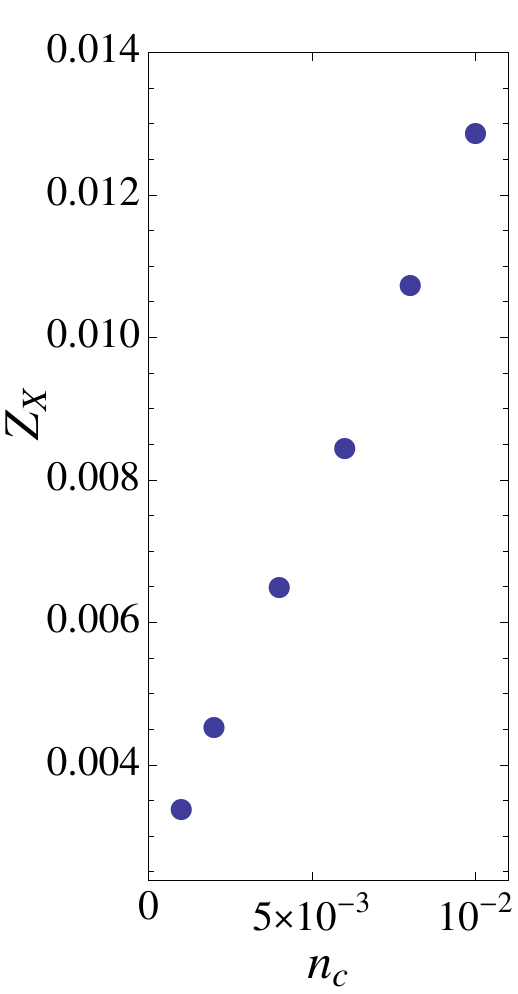}  
\caption{Left panel: lesser Green's function $-iG^{<}_{cc , 0}(\w)$ 
(in arbitrary units) for 
different densities of the conduction electrons $n_{c}$. Right panel: 
Dependence of the exciton weight $Z_{X}$ on $n_{c}$.}  
\label{specD}  
\end{figure} 
Let us start by 
assessing the accuracy of the two-particle correlation functions
in Eqs. (\ref{L<approx4}) and (\ref{L>approx4}). 
In Fig. \ref{Lless} we compare the numerical outcome of 
$L^{q,<}$ in Eq. (\ref{Lq<tot}) obtained by using the approximation of 
Eq. (\ref{L<approx4}) 
and the exact result of Eq. (\ref{Lexact}). 
The system parameters are $\mathcal{L}=80$, $\D=w/4 =1 \, \mathrm{eV}$, $T=4000 \, \mathrm{K}$ , 
$\m_{v}=2.35\, \mathrm{eV}$, $\m_{c}=2.65\, \mathrm{eV}$,  $U=2\, 
\mathrm{eV}$, $\eta=w/(4\mathcal{L})$. With these parameters the 
number of conduction electrons per unit cell  is 
$n_{c}=\frac{1}{\mathcal{L}}\sum_{k}f_{ck}\approx 10^{-2}$ and the 
solution of Eq.~(\ref{pseudoeigen}) for $q=0$ yields an exciton state 
 with binding 
energy $b_{X}\approx 0.42\, \mathrm{eV}$.
The accuracy of our approximation 
is excellent in the entire frequency domain. 
In particular both the exciton structure at $\approx 0.56\,\mathrm{eV}$ 
and the continuum of {\em eh} excitations above $\D = 1\,\mathrm{eV}$ 
are well reproduced; the relative error never exceeds 0.5\% and 
reaches its maximum at the 
exciton energy.

\begin{figure}[tbp]  
\includegraphics*[width=.4\textwidth]{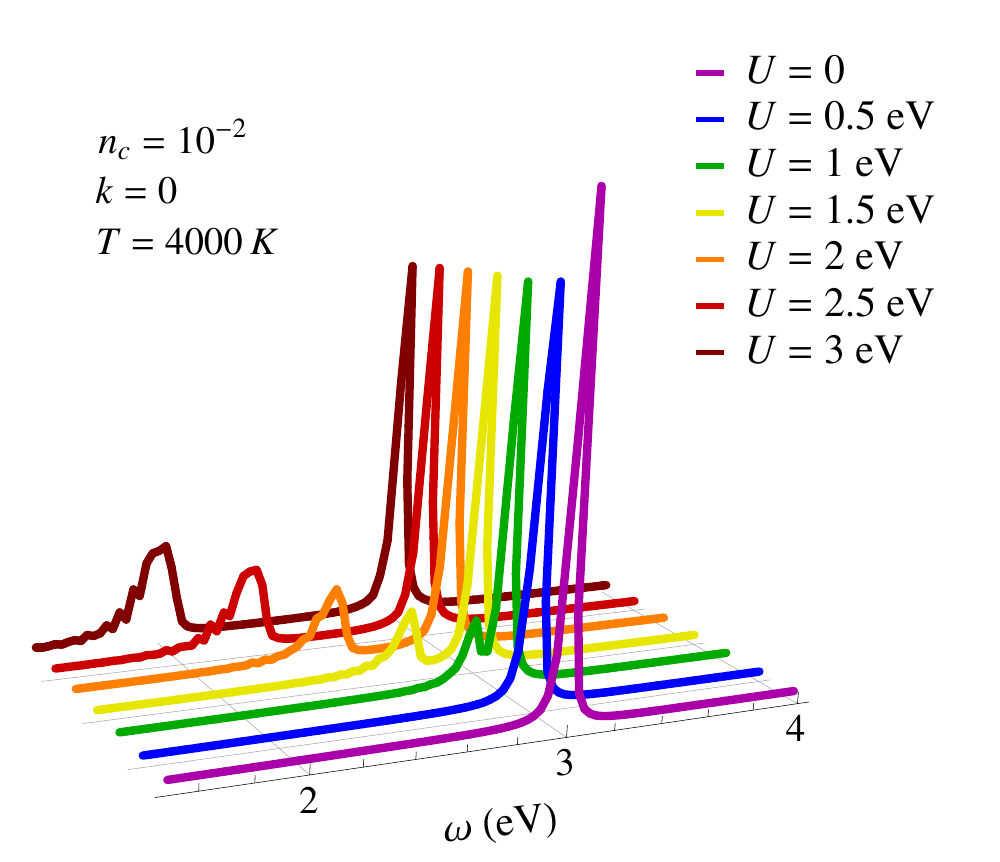}  
\includegraphics*[width=.4\textwidth]{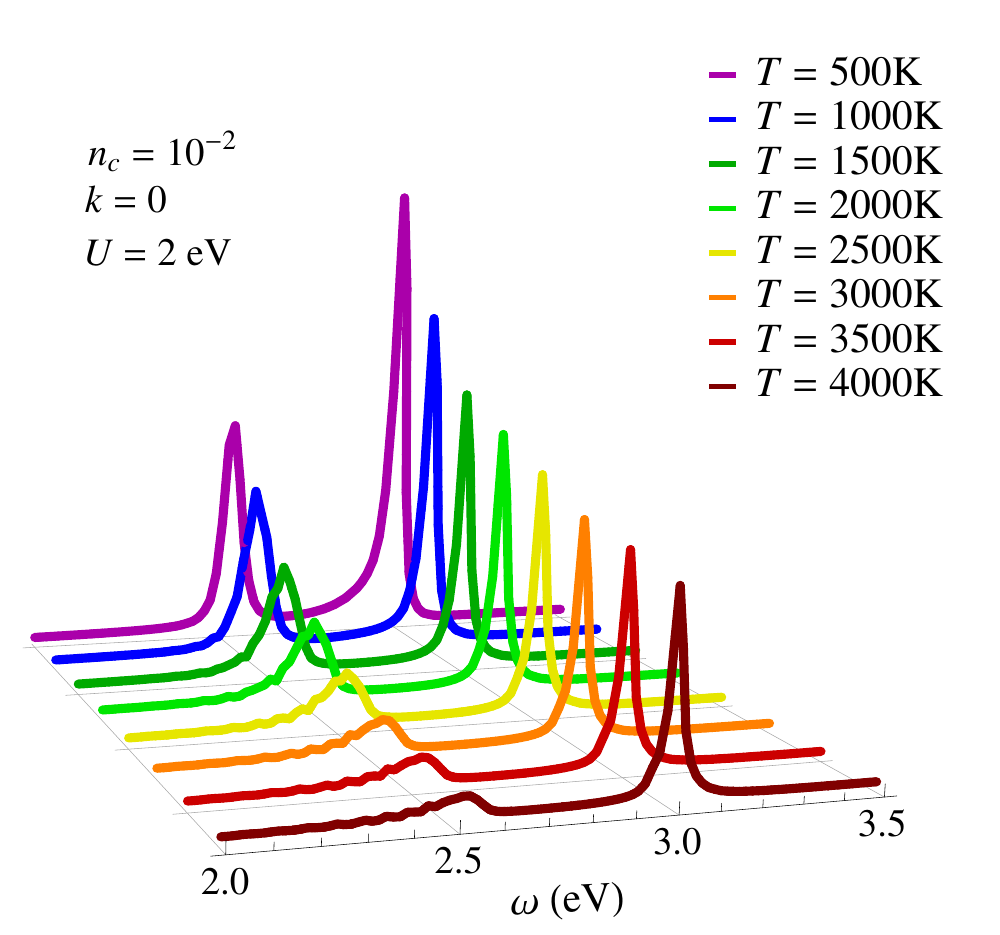}  
\caption{Lesser Green's function $-iG^{<}_{cc , 0}(\w)$ (in arbitrary 
units) for 
different interaction strength $U$ (top panel) and temperatures $T$ 
(bottom panel).}  
\label{specU}  
\end{figure} 
According to Eq.~(\ref{photocurr3}) 
the energy resolved photocurrent perpendicular to the surface
 is proportional to 
$G^{<}_{cc , 0}(\e-\w_{0})$.
In Fig. \ref{specD} (left panel) we show $G^{<}_{cc , 0}(\w)$ for different
carrier densities $n_{c}$.
At very low density $n_{c}\lesssim 10^{-4}$ the system is 
essentially in equilibrium and the photocurrent is vanishingly small   
(not shown).
At density  $n_{c} \approx 10^{-3}$ a {\em qp} peak at $\w 
\approx 3\, \mathrm{eV}$ appears. 
This corresponds to the 
removal energy of an excited electron 
from the bottom of the conduction band. This peak was 
absent for the singular occupation of the previous 
Section, i.e.,  $f_{ck}=\d_{k0}$, since in that case $T=0$.
At $n_{c} \approx 10^{-3}$ the exciton peak at  
$\e_{X}=\e_{c0}-b_{X} \approx 2.5\, \mathrm{eV}$ is still not visible 
because the exciton weight $Z_{X} = \int_{-\iif}^{\e_{c0}} 
\frac{d\w }{2\p}A(\w)$  is  still too small. The dependence of $Z_{X}$ 
on the density of conduction electrons is shown in the right panel 
of  Fig. \ref{specD} and it is by and large linear.
At higher density both the {\em qp} peak and the exciton peak 
become more pronounced. However, the latter acquires an {\it asymmetric} 
shape and an intrinsic {\it broadening}.
The broadening is not related to the lifetime of the exciton (which 
is infinite in our model) but origins from the 
fact that an electron with momentum $k$ participates to the 
formation of excitons of different total momentum. 
Of course the probability of finding an 
electron with $k=0$ in an exciton with total momentum $q$ decreases 
with increasing $|q|$ and hence with increasing the binding energy of the 
exciton. Thus the broadening is asymmetric and
proportional to the exciton bandwith.

In Fig. \ref{specU} we illustrate the evolution of $G^{<}_{cc , 0}(\w)$
by varying the interaction strength $U$ (top panel) and the 
effective temperature $T$  (bottom panel) at fixed density $n_{c}=10^{-2}$.
In the first case we clearly observe how the excitonic 
state develops. Starting from $U=0$ the exciton peak splits off 
from the {\em qp} peak and moves toward lower energies acquiring 
spectral weight and spreading over a finite energy window.
If we lower the temperature at fixed $U$ the bottom panel 
indicates that the exciton peak shrinks and raises. However, the spectral weight $Z_{X}$
remains essentially constant (not shown). This suggests that
the exciton peaks in TR-PE experiments should become more pronounced 
with increasing the delay between the pump and probe pulses
since the excited electron liquid in the conduction band (initially very hot) has more time to 
cool down before getting probed.

\begin{figure}[tbp]  
\includegraphics*[width=.46\textwidth]{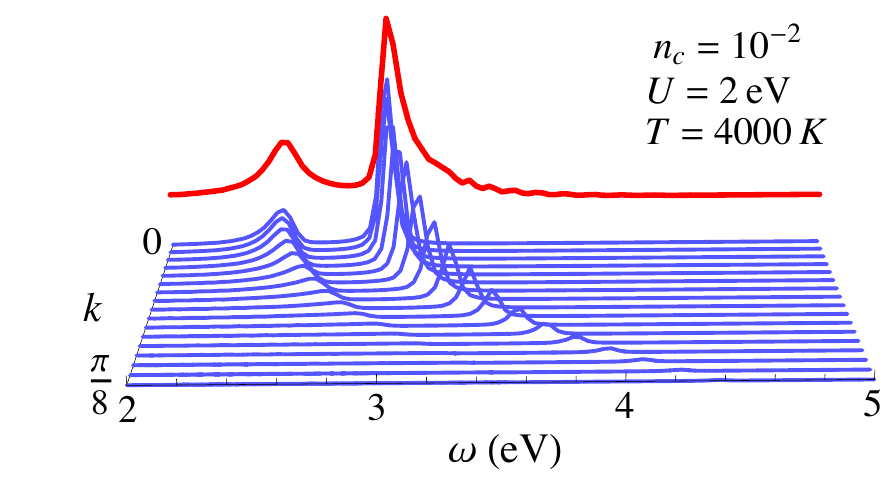}  
\caption{Lesser Green's function $-iG^{<}_{cc ,k}(\w)$ (in arbitrary 
units) for different 
momenta $k$ of the conduction electron. The (red) curve in the 
background is the integrated quantity $-i \int dk \,G^{<}_{cc ,k}(\w)$.}  
\label{specK}  
\end{figure} 

\begin{figure}[bp]  
\includegraphics*[width=.36\textwidth]{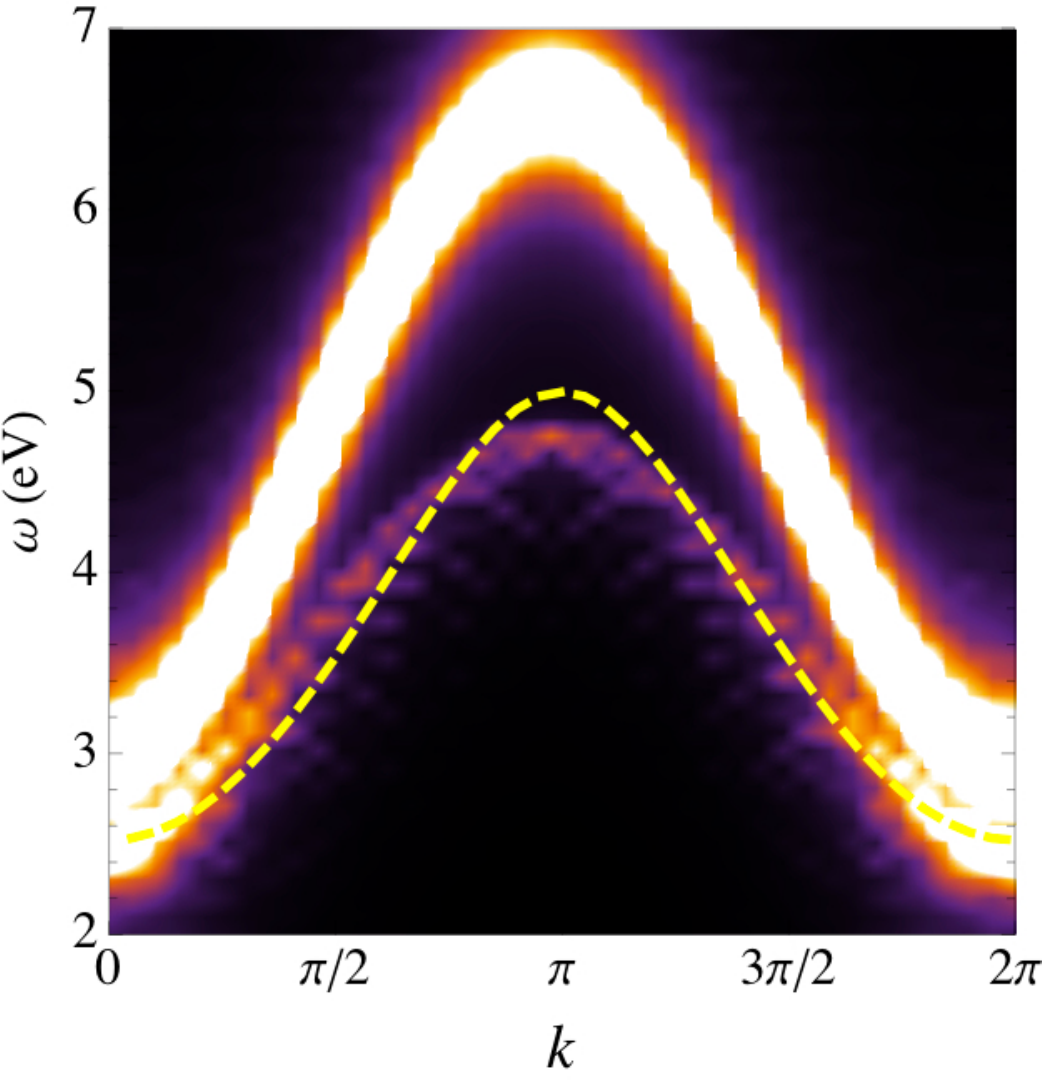}  
\caption{Momentum resolved and energy resolved excited spectral 
function $A_{k}(\w)$ in arbitrary units. The dashed line corresponds to the exciton 
dispersion of the system in equilibrium. }  
\label{ExcDisp}  
\end{figure} 

We have also calculated $G^{<}_{cc , k}(\w)$ for different momenta 
$k$ of the conduction electron. This quantity is 
relevant to address angle-resolved experiments.
In Fig.~\ref{specK} we plot $-iG^{<}_{cc , k}(\w)$ in the range $0<k<\p/8$. For 
$k>\p/8$ the lesser Green's function is strongly suppressed
by the Fermi function $f_{c}(\w)$, see Eq.~(\ref{heuristic6}).
It is interesting to observe that the angle-resolved  photocurrent 
gives, in principle, access to the dispersion of the {\em qp} bound in an 
exciton. 
In order to better appreciate this point we show in Fig.~\ref{ExcDisp}
the spectral function $A_{k}(\w) = 
i[G^{\rm R}_{cc,k}(\w)-G^{\rm A}_{cc,k}(\w)]$ for the same parameters 
of Fig.~\ref{specK}. From 
Eqs.~(\ref{mainsigma}) and (\ref{graexact}) we expect that the peaks in $A_{k}(\w)$ occur 
at the bare energy $\e_{ck}$ and at $\e_{v k-q}+\W^{X q}$ where $\l=X$
labels the energy needed to excite an exciton of momentum $q$.
In the quasi-stationary regime the residue 
$\bar{f}_{v k-q}\bar{F}^{\l q}+f_{v k-q}F^{\l q}$ of Eq.~(\ref{mainsigma}) is 
largest for $q\simeq k$ and hence the self-energy is dominated by the 
pole in $\e_{v0}+\W^{X k}$. The superimposed 
dashed line in Fig.~\ref{ExcDisp} corresponds to the value of 
$\e_{v0}+\W^{X k}$ as obtained from an 
{\em equilibrium} calculation.
More precisely we have solved 
Eq.~(\ref{pseudoeigen}) with equilibrium occupations
and then identified $\W^{X k}$ as  
the lowest (split-off) positive energy. If we write 
$\W^{X k}=\e_{ck}-\e_{v0}-b_{X,k}^{\mathrm{eq}}$ 
(where $\e_{ck}-\e_{v0}$ is the noninteracting excitation energy)
then $\e_{v0}+\W^{X 
k}=\e_{ck}-b_{X,k}^{\mathrm{eq}}$.
From Fig. \ref{ExcDisp} we see that $-iG^{<}_{cc , k}(\w)$ is peaked  
in $\e_{ck}$ and in 
the neighborhood of $\e_{ck}-b_{X,q}^{\mathrm{eq}}$, thus confirming 
the physical picture that the bare conduction electron splits into a dressed conduction 
{\em qp} and into a bound {\em qp}.  The discrepancy between the 
low-energy peak in $A_{k}(\w)$ and the 
equilibrium calculation (dashed line) is
due to the finite population of electrons in the conduction band. 
In general, the 
larger is $n_{c}$ and the more the bound {\em qp} dispersion differs from 
the one obtained by performing an equilibrium calculation. This 
points to the importance of solving the BSE with proper populations, 
as discussed in Section \ref{BSELsec}. It is worth noting that the 
bound {\em qp} dispersion depends on the band structure of 
the solid and can differ substantially from the one of 
Fig.~\ref{ExcDisp}. Nevertheless, our theory is not limited to the 
minimal model of Eq.~(\ref{minmodham}) and it can be applied to 
make predictions on real materials.

\section{Summary and Conclusions}
\label{conclusions-sec}

We developed a first-principles many-body diagrammatic approach to 
address TR and angle-resolved PE experiments in insulators and 
semiconductors with a low-energy spectrum dominated by exciton states. 
The time-dependent photocurrent can be 
calculated from a single-time convolution of the nonequilibrium lesser Green's 
function and embedding self-energy. The latter is independent of 
the interaction and it is completely determined by the shape of the 
probe pulse and by the dipole matrix elements. The calculation of the 
lesser Green's function does, in 
general, require the solution of the two-time Kadanoff-Baym 
equations.\cite{SvLbook,KBbook,KB.2000,DvL.2007,MSSvL.2009,BB.2013,PvFVA.2009,SBP.2016,SB.2016}
However, if we are interested in probing the excited 
system after the pumped electrons have reached 
a thermal distribution (in the conduction band) then a quasi-stationary picture applies. In 
this regime one can solve the simpler one-time Kadanoff-Baym equations 
for the populations and then use these populations as inputs for the many-body 
approach presented in this work. The take-home message is that 
excitonic features in TR-PE emerge provided that (1) the 
self-energy diagram contains the HSEX vertex
and (2) {\em excited qp} Green's function 
 are used to evaluate the self-energy diagrams.

The proposed theoretical framework has been applied to a minimal 
model Hamiltonian. We 
demonstrated that if the system is in a pure state with 
just one exciton then the many-body solution for the lesser Green's 
function coincides with the exact solution. At finite 
temperatures we studied several features of the  exciton peak. In 
addition to the intuitive red-shift with increasing the strength of 
the screened 
interaction we highlighted an asymmetric broadening which becomes 
more pronounced with increasing the density of electrons in the 
conduction band. We also showed that angle-resolved TR-PE 
spectroscopy can be 
used to calculate the bound {\em qp} dispersion and that this dispersion is 
in general different from the one obtained by solving the equilibrium 
BSE.

The proposed many-body approach is not the only first-principle 
method to tackle TR-PE spectra. Another popular method is Time-Dependent Density Functional 
Theory (TDDFT) which has already been applied to finite 
systems\cite{DGBCWR.2013,LDGR.2016,WDGCR.2016} and, as it was 
recently shown, could be used for solids as well.\cite{BRPE.2016} However, in 
practical applications TDDFT is implemented with local functionals of 
time and space and the resulting  spectrum is peaked at the 
Kohn-Sham  single particle energies. This is not always 
satisfactory and the only remedy consists in developing 
ultra-nonlocal functionals as discussed in 
Ref.~\onlinecite{USvL.2014}. Our work clearly shows that local functionals 
cannot describe exciton peaks in TR-PE.

Finally we wish to point out that a first-principle approach to TR-PE 
experiments is crucial for the correct physical interpretation of the 
behavior of the spectral features as the intensity and 
envelop of the pump field is varied. Our work represents a first 
step in this direction and paves the way toward a more general theory 
and numerical approach to access the far-from-relaxed regime of the 
system  during 
and shortly after the action of the pump.

\subsection*{Acknowledgements}

We acknowledge financial support by the Futuro in Ricerca Grant 
No. RBFR12SW0J of the Italian Ministry of Education, 
University and Research MIUR. G.S. and E.P. also acknowledge EC funding 
through the RISE Co-ExAN (GA644076). D.S. and A.M. also acknowledge 
funding  from the European Union project MaX Materials 
design at the eXascale H2020-EINFRA-2015-1, Grant Agreement No. 
676598 and Nanoscience Foundries and Fine Analysis - Europe 
H2020-INFRAIA-2014-2015, Grant Agreement No. 654360.


\begin{thebibliography}{10}
\bibitem{note}
The discussion can easily be generalized to situations where 
the system is left in an admixture of excited states.
\end{thebibliography}
\end{document}